\documentclass[11pt]{article}
\usepackage{graphicx}
\usepackage{amssymb}
\usepackage{amscd}
\usepackage{mathrsfs}
\usepackage{longtable,lscape}
\usepackage{amsthm}
\usepackage{amsfonts}
\usepackage{amsmath}
\usepackage{bbm}
\usepackage{float}
\usepackage{url}
\usepackage{hyperref}
\usepackage[margin=0.5cm,font=footnotesize]{caption}
\usepackage{lineno}
\usepackage[round]{natbib}
\usepackage{appendix}
\usepackage{subfig}

\begin{document}
\title{A Planck Radiation and Quantization Scheme \\ for Human Cognition and Language}
\author{Diederik Aerts and Lester Beltran  \vspace{0.5 cm} \\ 
        \normalsize\itshape
        Center Leo Apostel for Interdisciplinary Studies
        \\ 
        \normalsize\itshape
         Brussels Free University, Krijgskundestraat 33 \\ 
        \normalsize\itshape
         1160 Brussels, Belgium \\
        \normalsize
        E-Mails: \url{diraerts@vub.ac.be, diraerts@gmail.com},
           \\ \url{lbeltran@vub.ac.be, lestercc21@yahoo.com}
              	\\
              }
\date{}
\maketitle
\begin{abstract}
\noindent
As a result of the identification of `identity' and `indistinguishability' and strong experimental evidence for the presence of the associated Bose-Einstein statistics in human cognition and language, we argued in previous work for an extension of the research domain of quantum cognition. In addition to quantum complex vector spaces and quantum probability models, we showed that quantization itself, with words as quanta, is relevant and potentially important to human cognition. In the present work, we build on this result, and introduce a powerful radiation quantization scheme for human cognition. We show that the lack of independence of the Bose-Einstein statistics compared to the Maxwell-Boltzmann statistics can be explained by the presence of a `meaning dynamics', which causes words to be attracted to the same words. And so words clump together in the same states, a phenomenon well known for photons in the early years of quantum mechanics, leading to fierce disagreements between Planck and Einstein. Using a simple example, we introduce all the elements to get a better and detailed view of this `meaning dynamics', such as micro and macro states, and Maxwell-Boltzmann, Bose-Einstein and Fermi-Dirac numbers and weights, and compare this example and its graphs, with the radiation quantization scheme of a Winnie the Pooh story, also with its graphs. By connecting a concept directly to human experience, we show that entanglement is a necessity for preserving the `meaning dynamics' we identified, and it becomes clear in what way Fermi-Dirac addresses human memory. Within the human mind, as a crucial aspect of memory, in spaces with internal parameters, identical words can nevertheless be assigned different states and hence realize locally and contextually the necessary distinctiveness, structured by a Pauli exclusion principle, for human thought to thrive.
\end{abstract}

\section{Introduction}
Quantum cognition explores the possibility of using quantum structures to model aspects of human cognition \citep{aertsaerts1995,khrennikov1999,atmanspacher2002,gaboraaerts2002,bruzacole2005,busemeyeretal2006,bruzagabora2009,aerts2009a,lambertmoglianskizamirzwirn2009,aertssozzo2011,busemeyerbruza2012,havenkhrennikov2013,dallachiaraetal2015,melucci2015,pothosetal2015,blutnerbeimgraben2016,moreirawichert2016,broekaertetal2017,gaborakitto2017,surov2021}. Primarily, it is the structures of the complex vector space of quantum states and the quantum probability model that have been used fruitfully to describe aspects of human cognition. Recently, we have shown that quantum statistics, and more specifically the Bose-Einstein statistics, is also prominently and convincingly present in human cognition, and more specifically in the structure of human language \citep{aertsbeltran2020,aertsbeltran2022}. 
The presence of the Bose-Einstein statistics in quantum mechanics is associated with the `identity' and `indistinguishability' of quantum particles, and is probably one of the most still poorly understood aspects of quantum reality \citep{frenchredhead1988,saunders2003,mullerseevinck2009,krause2010,diekslubberdink2020}.
Although there are connections to entanglement, and in linear quantum optics there is now effective experimental use of the `indistinguishability' of photons to fabricate qubits, and thus `indistinguishability' is considered a `resource' for quantum computing, it remains one of the most mysterious quantum properties, also structurally different from entanglement \citep{francocompagno2018}. 
The original interest of one of us in identifying in human cognition and language an equivalent of this situation of `indistinguishability' in quantum mechanics, leading to the Bose-Einstein statistics, was motivated by working on a specific interpretation of quantum mechanics, called the `conceptuality interpretation' \citep{aerts2009b}. 
Thus, this original motivation was aimed more at increasing the understanding and explanation of what `identical' and `indistinguishable' quantum particles really are, rather than intended to introduce an additional rationale for research in quantum cognition. With a focus still primarily on this original motivation, work continued on the identification of a Bose-Einstein-like statistics by one of us, with a PhD student and collaborator, and more and better experimental evidence was collected for the superiority of Bose-Einstein statistics in modeling specific situations in human language as compared to Maxwell-Boltzmann statistics \citep{aertssozzoveloz2015b}. 
However, only by switching to a completely different approach, because the original approach of finding experimental evidence was not fully satisfying, was a layer of insight opened up, showing that the presence of the Bose-Einstein statistics in human cognition and language, makes a new unexplored structure of quantum mechanics important, and perhaps even crucial, to the research field of quantum cognition \citep{aertsbeltran2020,aertsbeltran2022}. 
The reason we dare to assert this, while being aware of the many aspects that remain to be explored, is because the `radiation and quantization scheme that we were able to identify as uniquely structurally underlying this presence of the Bose-Einstein statistics in human cognition and language' also appears to be the underlying structure of Zipf's law, and its continuous version, Pareto's law \citep{zipf1935,zipf1949,pareto1897}. Zipf's and Pareto's laws, which are empirical laws, without a theoretical foundation that is satisfactory, and about which there is a minimum of consensus, show up spontaneously in many different situations in many domains. These range from rankings of the size of cities \citep{gabaix1999}, rankings of the size of income \citep{aokimakoto2017}, but also rankings that seem much more marginal, such as the number of people who watch the same TV station \citep{erikssonetal2013}, or the rankings of notes in music \citep{zanette2006}, or the rankings of cells' transcriptomes \citep{lazardietal2021}. 
Thus, the radiation and quantization scheme that we propose in a detailed way in this article can be the foundation of these dynamical situations, where Zipf's or Pareto's law is empirically established. But the radiation and quantization scheme we propose is much richer that the simple ranking that gives rise to Zipf's law, although it contains and explains it, and introduces, for example, the notion of `energy level', and thereby the possibility of giving an energetic foundation to the dynamics underlying Zipf's and Pareto's laws. As is explicitly shown in the example we analyze, the notions of micro and macro states are introduced, and the associated entropy as a governing factor for this dynamics. This means that our approach can possess a broad application and value for very general and widely present dynamical situations in different domains of science, more specifically and among others in psychology and economics.

In Section \ref{boseeinstein} we outline the steps that took place in the identification of the Bose-Einstein statistics during the first phase, mainly also to show why it was difficult to find enough experimental evidence to push that approach through to a satisfactory conclusion. In Section \ref{radiationlawhumanlanguage}, we analyze the Planck and Wien radiation laws, which were already previously an inspiration to us, but never directly. We explain how a direct application of the continuous functions, Bose-Einstein and Maxwell-Boltzmann, led to a totally different, and much more powerful approach in terms of identifying the Bose-Einstein statistics in human cognition and language. We outline in what way the historical discussions that took place, between Planck, Einstein, and Ehrenfest, before Einstein made the prediction of the possibility of a Bose-Einstein condensate for a boson gas in 1925, inspired and helped us to properly develop the radiation quantization scheme, the elaboration of which is the subject of this article. 
We illustrate the radiation and quantization scheme using the example story of Winnie the Pooh, the first story we analyzed in \citet{aertsbeltran2020}, and illustrate the radiation quantization scheme by means of the graphs of this Winnie the Pooh story. In Section \ref{dynamicsofmeaning}, we then further develop this radiation and quantization scheme. We identify the lack of independence embedded in the Bose-Einstein statistical probabilities, as clearly and sharply put forward by Einstein and Ehrenfest for the radiation and quantization scheme of light, as caused by a `dynamics of meaning' intrinsic to human creations by what humans and their creations are. We analyze this `dynamics of meaning' in detail using an example of four words distributed over seven energy levels with total energy equal to seven. Using the example, we show how micro and macro states are formed, and how the exact combinatorial-based Maxwell-Boltzmann, Bose-Einstein, and Fermi-Dirac functions, are calculated. We set up the graphs of this example, and compare them to the graphs of the Winnie the Pooh story. We elaborate on how to introduce `that which radiates' into the scheme, and explain how in a natural way this must possess the form of a memory structure, more specifically the human memory in the case of human cognition and language. 
We analyze, by grounding concepts in human experience, how entanglement is indispensable when texts seek precision of formulation and maximization of meaning. The bosonic and fermionic structures can also be understood from this grounding, the bosonic present as the basic element of communication, thus of language, and the fermionic present as the basic element of `that which radiates', `the speaker', `the writer', in short `the memory'. 

\section{Identity and indistinguishability}
\label{boseeinstein}
Our original intuition, and what set us on the track to identify `identity' and `indistinguishability' in human cognition, and more specifically, in human language, was the following. If in a story the expression `eight cats' comes up, then generally these cats are completely `identical', and `indistinguishable' from each other within that story, and within the way we use the expression `eight cats' in human language. 
At least, when we consider the story in its generality, such that it can be read and/or listened to by any human mind linguistically capable of reading a story and/or listening to it -- this nuance is important when we introduce the difference between bosonic and fermionic structures in Section \ref{meaningdynamics}. However, `identity' and `indistinguishability' for quantum particles in quantum mechanics is not simply an epistemic matter, but also involves two very different statistical behaviors, compared to how classical particles behave. More specifically, the Maxwell-Boltzmann statistics, which describes the behavior of classical particles, is replaced by, on the one hand, the Bose-Einstein statistics when the quantum particles are bosons, and on the other hand, the Fermi-Dirac statistics when the quantum particles are fermions. Bosons are the particles of the quantum forces, electromagnetism and its photons, the weak force and its $W$ and $Z$ bosons, and the strong force and its gluons, and bosons are further characterized by carrying a spin of integer value, i.e. spin 0, spin1, spin 2, etc. Fermions on the other hand are the particles of matter, quarks, electron, muon, tauon and their respective neutrinos, and they are further characterized by carrying a spin of half integer value, spin 1/2, spin 3/2, spin 5/2, etc. For the replacement in human language of the Maxwell-Boltzmann statistics by a statistics that would be closer to Bose-Einstein, we were initially guided by a specific reasoning, before we began a systematic search for evidence for it. Let us illustrate this reasoning with an example. 

Suppose we go to a farm where there are about as many cats available as dogs, and we ask the farmer to choose two animals for us, where each animal is either a cat or a dog. And let us suppose that the farmer chooses for us, with probability $1/2$ to choose a cat, and probability $1/2$ to choose a dog. We also make the hypothesis that the farmer's choice of the first animal does not affect the farmer's choice of the second animal, e.g. asking the farmer to choose at random based on the toss of a non-biased coin. 
In that case, it is easy to check that we have probability $1/4=1/2 \cdot 1/2$ that we are offered by the farmer two cats, probability $1/4=1/2 \cdot 1/2$ that we are offered two dogs, and probability $1/2=1/2 \cdot 1/2+1/2 \cdot 1/2$ that we are offered a cat and a dog. This is a situation well described using the Maxwell-Boltzmann statistics, and indeed, the way these probabilities form, as products of independent individual probabilities for the animals separately, is typical of the Maxwell-Boltzmann statistics. 
We obtain a Maxwell-Boltzmann description in its optimal detail if we consider the farmer's choice of `a cat and a dog' different from the choice of `a dog and a cat'. For example, if the cats and dogs were put in two baskets, and we can easily distinguish the two baskets, and hence also the choices `cat in the first basket and dog in the second basket' is easy to distinguish from `dog in the first basket and cat in the second basket'. The probabilities then become $1/4$, $1/4$, $1/4$ and $1/4$, each being the product of the independent probabilities for cat or dog, $1/2$ and $1/2$. 
In the real situation of the example, the difference between the farmer offering `a cat and a dog' or `a dog and a cat' is not relevant, and hence we arrived at $1/4$, $1/4$ and $1/2$ for the respective probabilities, and we will call this a `Maxwell-Boltzmann description with presence of epistemic indistinguishability' between `cat and dog' and `dog and cat', the underlying statistics indeed remains Maxwell-Boltzmann. However, the Bose-Einstein statistics intrinsically deviates from this Maxwell-Boltzmann with epistemic indistinguishability. Indeed, Bose-Einstein assigns $1/3$, $1/3$, and $1/3$ probabilities to `two cats', `two dogs', and `a cat and a dog', respectively. Let us develop further our initial reasoning on this issue to show how we saw a way to support the presence of the Bose-Einstein probabilities in human cognition. Indeed, our initial reasoning argued that, although they clearly cannot arise from products of probabilities of the farmer's independent separate choices, they may nevertheless be understood if we consider human cognition and human language more closely. Let us suppose that it is not the farmer who chooses, but that at home the children, and there are two of them, a son and a daughter, are doing the choosing. And the three choices are presented to them like this, what do you wish, two kittens, two puppies, or a kitten and a puppy. All sorts of factors can play into how such a choice turns out, but there seems no reason to assume that it will turn out to be a Maxwell-Boltzmann choice with epistemic indistinguishability between `a kitten and a puppy' and `a puppy and a kitten', and hence with probabilities $1/4$, $1/4$ and $1/2$ for the choices of the three possibilities, `two kittens', `two puppies' or `a kitten and a puppy'. On the contrary, it may well be the case that `a cat and a dog' as a choice is rather problematically presented to the children by the parents, and that this alternative is even assigned less than $1/3$ probability. On the other hand, suppose the children do not agree, and it is decided to let them choose separately and independently from each other, then Maxwell-Boltzmann with epistemic indistinguishability and hence probabilities $1/4$, $1/4$ and $1/2$ comes back into the picture.

This `indistinguishability', of concepts, along with `lack of independence', of corresponding probabilities, and how it possibly would appear in human cognition and language, as illustrated by the example of `cat' and `dog' above, was pondered about by one of the authors in the 2000s, during the emergence of the research domains of quantum cognition and quantum information retrieval \citep{aertsaerts1995,khrennikov1999,atmanspacher2002,aertsczachor2004,vanrijsbergen2004,widdows2004,bruzacole2005,busemeyeretal2006}. A first attempt to find experimental evidence for the idea took place in the years that followed by using how aspects of human cognition could be investigated starting from customized searches on the World-Wide Web. More specifically, combinations of quantities of cats and dogs were considered, in varying sizes, such as `eleven cats', `ten cats and one dog', `nine cats and two dogs', `eight cats and three dogs', and so on, up to `two cats and nine dogs', `one cat and ten dogs', ending with `eleven dogs'. So the combinations were chosen in such a way that the sum of the number of cats and dogs always equals eleven. If we think back to the farmer who lives on a farm with many cats and dogs who would be asked to fill eleven baskets, in each basket a cat or a dog, again it is the Maxwell-Boltzmann distribution that comes up. The situation of `five cats and six dogs' will be much more common than the situation of `eleven cats', because each basket will be filled in an independent manner with either a cat or a dog, with probability $1/2$ for this choice, bearing in mind that there are as many cats as dogs on the farm, and the farmer is assumed to choose in an independent manner for each basket again. The idea was to search the frequency of occurrence of these twelve sentences with varying combinations of numbers of cats and dogs on the World-Wide Web using the Google search engine. The underlying idea was identical to that expressed in the example above, namely, that by the human mind is chosen differently than the way cats and dogs end up in baskets by the farmer's choices. And more specifically, that the Bose-Einstein statistics might provide a better model for this human way of choosing than this is the case for the Maxwell-Boltzmann statistics. The results indeed showed that the Maxwell-Boltzmann statistics is not respected by the frequencies of occurrence of these sentences on the World-Wide Web, and that the Bose-Einstein statistics may well provide a better model (see Section 4.3 of \citet{aerts2009b}). The results were published in the year 2009 which, after a very active second half of the 2000s, with several proceedings of conferences, also spawned the first comprehensive special issue about `Quantum Cognition' \citep{bruzagabora2009}. 
 It took several years before Tomas Veloz, in the work for his PhD, undertook a much more systematic study, starting from the same idea, but now adding other pairs of concepts than the `cat, dog', such as `sister, brother', `son, daughter', etc, and also considering other numbers than only eleven for the sums to be equal to. There was also a cognitive experiment set up, where participants were asked to choose between several of the proposed alternatives. This time, it became clear in a much more convincing way that Bose-Einstein provided a better model compared to Maxwell-Boltzman for the obtained outcomes of the considered cases \citep{aertssozzoveloz2015b}. 

Both in \citet{aerts2009b} and in \citet{aertssozzoveloz2015b}, regarding the evidence gathered from searches on the World-Wide Web, there was a weak element, namely that search engines such as Google or Yahoo do not make a very reliable count of the number of web pages. While these numbers of web pages played an important role in calculating the probabilities, and thus in comparing Bose-Einstein statistics with Maxwell-Boltzmann statistics as a model for the results obtained. The idea, then, was to address the relative weakness of previous attempts to find experimental evidence by replacing searches on search engines such as Google and Yahoo with searches in corpuses of documents. More specifically, the use of the corpuses, `Google Books' (\url{https ://googl ebook s.byu.edu/x.asp}), `News on Web' (NOW) (\url{https ://corpu s.byu.edu/now/}), and `Corpus of Contemporary American English' (COCA) (\url{https ://corpu s.byu.edu/coca/}), was investigated. With these searches on these corpuses of documents one was assured of the exactness of the frequencies of occurrences indicated with a search, so that the probabilities calculated from them carried great experimental reliability. The corpuses proved perfectly suited to study entanglement in human cognition and language \citep{beltrangeriente2019}. However, related to Maxwell-Boltzmann and Bose-Einstein, the presence of the sentences needed to test the difference between the two statistics was found to be too small in even Google Books, the largest of the three corpuses. Most searches of these sentences led to zero hits which was mainly also due to the nonconventional nature of their content, as there are not many stories where, for example, `seven cats and four dogs' is mentioned. When it became clear that Google Books is also the largest corpus available anyway, we knew that only the World-Wide Web is large enough to demonstrate `in this way' the superior modeling power in human cognition of the Bose-Einstein statistics over the Maxwell-Boltzmann statistics. The PhD student among the authors of the present article then tried to use Google anyway, where he decided to count the hits manually. This was possible because, with the current size of the World-Wide Web, these `numbers of hits' for the sentences such as `seven cats and four dogs' came to between 100 and 250, which made it possible to count them manually and thus to arrive at a reliable result with respect to the frequencies of occurrence. 
In an even more convincing way than this was the case in previous studies, the Bose-Einstein distribution proved to yield a better model for the collected data than this was the case for the Maxwell-Boltzmann distribution \citep{beltran2021}. At this stage, however, where it proved impossible to test the original idea in even the largest of the available corpus of documents, namely Google Books, we began to look for possible other ways to know the nature of the underlying statistics of human cognition and language. And it turned out that indeed a much more direct experimental test was possible than the ones we had tried so far. 

\section{A radiation and quantization scheme for human cognition}
\label{radiationlawhumanlanguage}
During the reflections set forth in Section \ref{boseeinstein}, and being in pursuit of the Bose-Einstein statistics in human cognition and language, the closest example of a Bose-Einstein statistics in physical reality, was also always a subject of our reflection. This closest example is that of electromagnetic radiation, light, consisting of photons, which are bosons, and therefore behave following the Bose-Einstein statistics. There is also a historically very interesting aspect to this example of `light with its photons', for it is the study of how a material entity that is heated begins to emit light radiation that led scientists on the path to quantum mechanics. Let us, write down the radiation law for light, for its form has played a role in our further investigation of the presence of the Bose-Einstein statistics in human cognition and language.
\begin{eqnarray} \label{planckradiation}
B(\nu,T) = {2 h \nu^3 \over c^2} {1 \over {e^{h\nu \over kT} - 1}}
\end{eqnarray} 
The quantity $B(\nu,T)$ is the amount of energy, per unit surface area, per unit time, per unit solid angle, per unit frequency at frequency $\nu$ that is radiated by a material entity that has been heated. The temperature T is measured in degrees Kelvin, hence with reference to the absolute zero. Three constants occur in this radiation law for light, $h$ is Planck's constant, $c$ is the speed of light, and $k$ is Boltzmann's constant. 
We digress a little further, also on some of the historical aspects of how this radiative law of light played a role in the emergence of quantum mechanics, because that's also how we came to the much more direct and powerful way of identifying the Bose-Einstein statistics in human cognition and language. 

\begin{figure}[h!]
\begin{center}
\includegraphics[width=12cm]{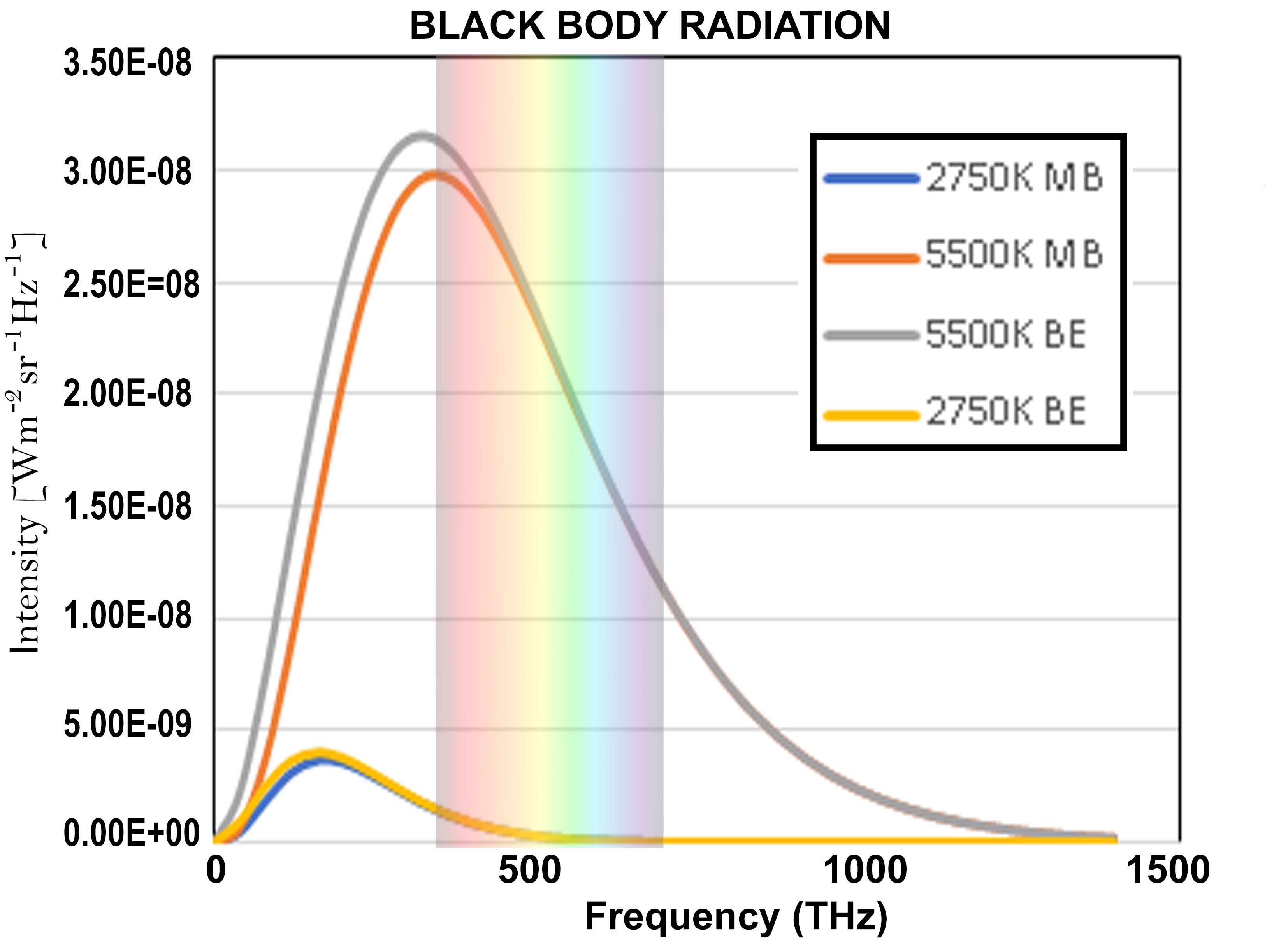}
\end{center}
\caption{The gray and yellow graphs are the Planck's radiation law curves for $5500^\circ$ Kelvin, which is approximately the temperature of the sun, and $2750^\circ$ Kelvin, respectively, and the red and blue graphs are the Wien's radiation law curves for $5500^\circ$ Kelvin and $2750^\circ$ Kelvin, respectively. We can clearly see on the graphs that for high frequencies, starting from red light, the two curves, the Wien curve and the Planck curve, practically coincide, and this was the reason that experiments were needed with low frequency electromagnetic waves to measure the difference between the two.}
\label{BlackBodychartfigure}
\end{figure}

It was Max Planck who in 1900 proposed the function shown in formula (\ref{planckradiation}) as the function that describes the radiation from a black body and thus initiated the first phase (1900-1925) of the development of quantum physics \citep{planck1900}. In physics, a `black body' means, `a body that absorbs all incident light'. The radiation law formulated by Planck was thus supposed to describe the radiation emitted by a material entity as a result of heating, without involving reflected radiation. Often the theory developed during these years, which was primarily a collection of novel but mostly ad hoc rules, is called the `Old Quantum Theory'. Planck began the study of the radiation from a black body using a law formulated by Wilhelm Wien, which showed very good agreement with the experimental measurements available at the time \citep{wien1897}. During the year 1900, experiments were carried out with long wavelength radiation which showed indisputably that Wien's law failed for this type of radiation \citep{rubenskurlbaum1900}, while Planck was then fully engaged in proving a thorough derivation from Boltzmann's thermodynamics for Wien's law. The introduction of the constant $h$, now named after him, together with the constant $k$, which Planck named after Boltzmann, was clearly intended primarily as a rescue operation for his thoroughly elaborated thermodynamic theory of electromagnetic radiation. Indeed, in an almost miraculous way, the new law now also described very well the long wavelength radiation that had discredited Wien's original law.  
A study of the scientific literature and correspondences between scientists of the period also makes it clear that Planck was not thinking about `quantization' at all, in the sense we understand it today. It was therefore Albert Einstein, in his article describing the photoelectric effect \citep{einstein1905}, who was the first to consider the $h\nu$ introduced by Planck, as explicit `quanta of light', later called `photons' \citep{klein1961}. Let us consider explicitly what that law of Wien is, and as will become clear, it is relevant to the part of our inquiry that we wish to bring forward here. We will write Wien's law in the following form
\begin{eqnarray} \label{wienradiation}
B(\nu,T)_{Wien} = {2 h \nu^3 \over c^2} {1 \over {e^{h\nu \over kT}}}
\end{eqnarray}
where it can be considered an approximation of Planck's law. Indeed, if $h\nu >> kT$, which is the case for short wave length radiation, we can approximate $1 / (e^{h\nu \over kT} - 1)$ well by ${1 / {e^{h\nu \over kT}}}$), and then Planck's law reduces to Wien's law.
Historically, it was of course not Wien who introduced the two constants, $h$ and $k$, that was Planck, Wien proposed an arbitrary constant in the law as he formulated it. In Figure \ref{BlackBodychartfigure} we have represented both Planck's radiation law formula and Wien's approximation. The gray and yellow graphs are Planck's radiation law curves for $5500^\circ$ Kelvin, which is approximately the temperature of the sun, and $2750^\circ$ Kelvin, respectively, and the red and blue graphs are the Wien's radiation law curves for $5500^\circ$ Kelvin and $2750^\circ$ Kelvin, respectively. 
We can clearly see on the graphs that for high frequencies the two curves, the Wien curve and the Planck curve, practically coincide, and this was the reason that experiments were needed with low frequency electromagnetic waves to measure the difference between the two \citep{rubenskurlbaum1900}.
 
What we want to make clear now is that the portions of the functions in the two laws that are different are exactly the signatures of the two statistics we talked about in Section \ref{boseeinstein}, the Bose-Einstein and the Maxwell-Boltzmann. Or, to put it more precisely,
\begin{eqnarray} \label{boseeinsteindistribution}
N(E_i) = {1 \over {Ae^{{E_i \over B}}-1}}
\end{eqnarray}
and
\begin{eqnarray} \label{maxwellboltzmanndistribution}
N(E_i) = {1 \over Ce^{{E_i \over D}}}
\end{eqnarray}
the two portions of the laws that are different, are functions that represent the continuous form respectively of the Bose-Einstein statistics, and of the Maxwell-Boltzmann statistics. We have this time replaced the term $h\nu$ with $E_i$, with an index $i$, which stands for `energy level number $i$', and $N(E_i)$ is the number of quanta which has energy $E_i$. We introduced two general constants, $A$ and $B$ for the Bose-Einstein distribution, and $C$ and $D$ for the Maxwell-Boltzmann distribution. These two constants will be completely determined by summing the $i$ over all energy levels, from $0$ to $n$, where then $N$ is the total number of particles, and $E$ is the total energy. Hence
\begin{eqnarray} \label{BEconstraint}
\sum_{i=0}^n {1 \over {Ae^{{E_i \over B}}-1}} = N \quad {\rm and} \quad \sum_{i=0}^n {E_i \over {Ae^{{E_i \over B}}-1}} = E
\end{eqnarray}
for the case of Bose-Einstein, and
\begin{eqnarray} \label{MBconstraint}
\sum_{i=0}^n {1 \over Ce^{{E_i \over D}}} = N \quad {\rm and} \quad \sum_{i=0}^n {E_i \over Ce^{{E_i \over D}}} = E
\end{eqnarray}
for the case of Maxwell-Boltzmann.

As we mention in Section \ref{boseeinstein}, as a consequence of even the largest available corpus of documents being too small in to allow the collection of relevant hits for sentences such as `four cats and seven dogs', we had begun to look for possible other ways to identify the statistical structure in human cognition and language. And, at some point, the idea arose to use the continuous functions (\ref{boseeinsteindistribution}) and (\ref{maxwellboltzmanndistribution}) that are the fingerprints of Bose-Einstein or Maxwell-Boltzmann to identify the underlying statistics in human cognition and language.This implied introducing `energy levels' for words, with `the number of times a word occurs in a text' as a criterion for the `energy level' the word is in within the text. Indeed, we were inspired by Planck's and Wien's radiation laws for a black body, which indeed describe `the number of photons' emitted over the different energy levels. Both laws also classify the photons by energy level, given by Planck's formula $E=h\nu$ which links the energy level $E$ to the frequency $\nu$ of the radiation, as can be seen in Figure \ref{BlackBodychartfigure}. The frequency ranges from zero, hence energy level equal to zero, to ever higher frequencies, hence increasing energy levels, with fading radiation for increasing frequencies. 

Note also that both (\ref{boseeinsteindistribution}) and (\ref{maxwellboltzmanndistribution}) have only two constants in them, $A$ and $B$, and $C$ and $D$, respectively, which will be determined by the total number of words of the text, $N$, and by the total energy, $E$, of all the radiation generated by the text. And this total number of words and this total energy of radiation is uniquely determined by (\ref{BEconstraint}) and (\ref{MBconstraint}), respectively, in the Bose-Einstein case or the Maxwell-Boltzmann case. This means that there are `no parameters' to possibly make a Bose-Einstein model fit better than a Maxwell-Boltzmann model, or even to fit either one. We were not fully aware of this total lack of any possible `fitting parameters', but as we began to pick a text to test our idea for the first time, this insight dawned. And so we knew that even with this first test, with the first text we would pick, it would be all or nothing. And at worst, neither the Bose-Einstein, or the Planck formula, nor the Maxwell-Boltzmann, or the Wien formula, would show a fitting of any quality. That would then indicate that the underlying statistics of human cognition and language are of a completely different nature, with the rather inevitable result that our basic idea was just plain wrong. We chose a text from the Winnie the Pooh stories, and more specifically the story titled `In Which Piglet Meets a Haffalump' \citep{milne1926}. We counted the words by hand, to arrive at a ranking starting from energy level $E_0=0$, for the word that occurs the most, which was the word {\it And} -- we will notate `words' with capital letter and in italic, as in our previous articles.Then energy level $E_1=1$ for the word that occurs the second most, and that was the word {\it He}, and then so on, ..., up to the words that occur only once, such as the one we classified as the last, the word {\it You've}. In Table \ref{piglethaffalunmp} we have presented a portion of these ranked words, the full table is provided with the supplementary material. 

In the first column of Table \ref{piglethaffalunmp}, the words are ranked, starting with the most occurring word, {\it And}, to the least occurring word, {\it You've}. The Winnie the Pooh story we selected contained $543$ different words, and so that means that the story, within our scheme, is assigned $543$ different energy levels, starting from energy level equal to $0$, to energy level equal to $542$. 
\begin{table}[h!]
\centering
\caption{An energy scale representation of the words of the Winnie the Pooh story `In Which Piglet Meets a Haffalump' by A. A. Milne as published in \citet{milne1926}. The words are in the column `Words' and the energy levels are in the column `$E_i$', and are attributed according to the `numbers of appearances' in the column `$N(E_i)$', such that  lower energy levels correspond to a higher order of appearances. The `amounts of energies radiated by the words of energy level $E_i$' are in the column `$E(E_i)$'. In the columns `B-E', `M-B', `Energ B-E', and `Energ M-B' are respectively the predicted values of the Bose-Einstein and the Maxwell-Boltzmann model of the `numbers of appearances', and of the `radiated energies'. In the graphs of Figure \ref{piglethaffalunmpenergygraphfigure}, we can see that a maximum is reached for the energy level $E_{71}$, corresponding to the word {\it First}, which appears seven times in the Winnie the Pooh story. If we use the analogy with light, we can say that the radiation spectrum of the story `In Which Piglet Meets a Haffalump' has a maximum at {\it First}, which would hence be, again in analogy with light, the dominant color of the story. We have indicated this radiation peak by underlining the word {\it First}, its energy level $E_{71}$, and the amount of energy 522.79 the story radiates, following the Bose-Einstein model. The complete table can be found with the supplementary material.}
\label{piglethaffalunmp}
\begin{tabular}{p{1.2cm}p{1.2cm}p{1.2cm}p{1.7cm}p{1.7cm}p{1.2cm}p{2cm}p{2cm}}
\hline
  Words  &  $E_i$ &  $N(E_i)$ & B-E mod &  M-B mod & $E(E_i)$ & Energ B-E & Energ M-B  \\
\hline
          {\it And}  & 0 & 133 & 129.05 & 28.29 & 0    & 0     & 0  \\
         {\it He} & 1 & 111 & 105.84 & 28.00 & 111 & 105.84 & 28.00  \\
        {\it The} & 2 & 91  & 89.68   & 27.69 & 182 & 179.36 & 55.38  \\
            {\it It} & 3 & 85  & 77.79   & 27.40 & 255 & 233.36 & 82.19  \\
            {\it A} & 4 & 70  & 68.66   & 27.11 & 280 & 274.65 & 108.43  \\
                       {\it To} & 5 & 69  & 61.45   & 26.82 & 345 & 307.23 & 234.09  \\
        {\it Said} & 6 & 61  & 55.59   & 26.53 & 366 & 333.55 & 159.20  \\
        {\it Was} & 7 & 59  & 50.75   & 26.25 & 413 & 355.24 & 183.76  \\
      {\it Piglet} & 8 & 47  & 46.68   & 25.97 & 376 & 373.40 & 207.78  \\
              {\it I} & 9 & 46  & 43.20   & 25.70 & 414 & 388.82 & 231.27  \\
        {\it That} & 10 & 41 & 40.21  & 25.42 & 410 & 402.05 & 254.24  \\
       {\it Pooh} & 11 & 40 & 37.59  & 25.15 & 440 & 413.52 & 276.69  \\
         \vdots &  \vdots &  \vdots   &  \vdots    & \vdots &  \vdots &  \vdots &  \vdots  \\
          {\it Did} & 70 & 7   & 7.47    & 13.40 & 490 & 522.78 & 937.67  \\
 {\it \underline {First}} & {\bf \underline {71}} & 7   & 7.36    & 13.25 & 497 & {\bf \underline{522.79}} & 940.96  \\ 
                  {\it Have} & 72 & 7   & 7.26    & 13.11 & 504 & 522.78 & 944.08  \\ 
         \vdots &  \vdots &  \vdots   &  \vdots    & \vdots &  \vdots &  \vdots &  \vdots  \\
                       {\it Word} & 539 & 1   & 0.67    & 0.09 & 539 & 359.58 & 48.22  \\
                   {\it Worse} & 540 & 1   & 0.67    & 0.09 & 540 & 359.24 & 47.80  \\
                      {\it Year} & 541 & 1   & 0.66    & 0.09 & 541 & 358.90 & 47.38  \\
                   {\it You've} & 542 & 1   & 0.66    & 0.09 & 542 & 358.55 & 46.96  \\
                   \hline
                                     &      &2655   &2655.00    &2654.96 & 242{,}891 & 242{,}891.01 & 242{,}889.76  \\
\end{tabular}
\end{table}
In the second column of Figure \ref{piglethaffalunmp}, we have represented these energy levels, i.e. running from the lowest energy level, $E_0 = 0$, to the highest energy level $E_{542} = 542$, achieved in this particular story of Winnie the Pooh. The third column shows `the number of times' $N(E_i)$ the with energy level $E_i$ corresponding word in the first column occurs in the Winnie the Pooh story. Specifically, the word {\it He}, corresponding with energy level $E_1$, occurs $111$ times, and the word {\it First}, corresponding with energy level $E_{71}$, occurs $7$ times. The total amount of words of the Winnie the Pooh story is equal to $2655$, and, per construction, is therefore equal to $N = \sum_{i=0}^{542} N(E_i)$. 
Before we explain columns four and five, we want to tell what is in column six, because together with the first three columns, we have then described all the information that has its roots in the experimental data collected from the Winnie the Pooh story. The quantities in column six are the `energies per energy level' $E(Ei)$. More specifically, we multiply the number of words that are in energy level $E_i$, i.e. $N(E_i)$, by the value $E_i$ of this energy level, which gives us $E(E_i)=N(E_i)E_i$, namely the energy radiated from that specific energy level $E_i$, where each word radiates and contributes to that total energy $E(E_i)$ per energy level $E_i$. Once we have entered that `energy per energy level' in column six, we can immediately calculate the total energy $E$ radiated by the entire text, by taking the sum of all these energies per energy level, so $E = \sum_{i=0}^{542}E(E_i) = \sum_{i=0}^{542}N(E_i)E_i$.  
\begin{figure}[h!]
\begin{center}
\includegraphics[width=10cm]{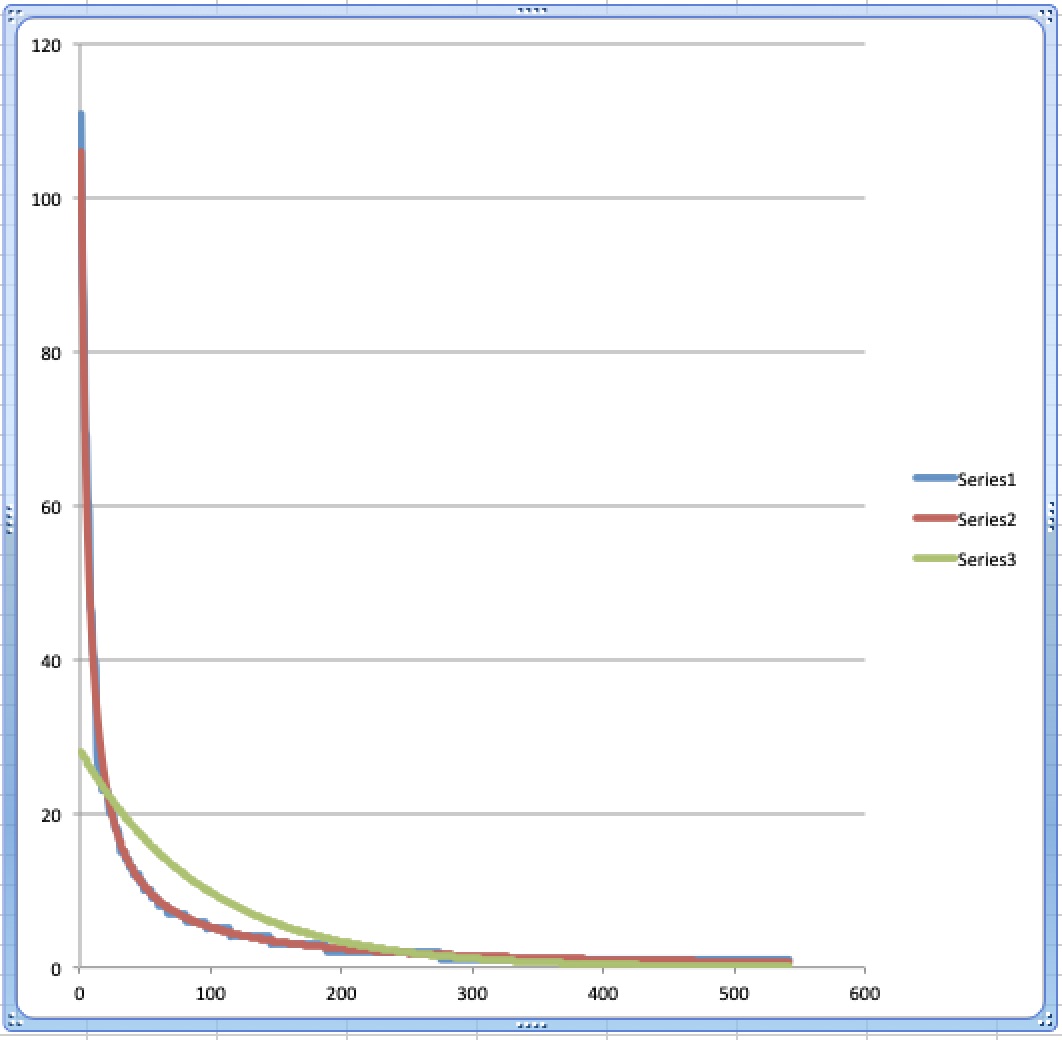}
\end{center}
\caption{A representation of the `number of appearances' of words in the Winnie the Pooh story `In Which Piglet Meets a Haffalump' \citep{milne1926}, ranked from lowest energy level, corresponding to the most often appearing word, to highest energy level, corresponding to the least often appearing word as listed in Table \ref{piglethaffalunmp}. The blue graph (Series 1) represents the data, i.e. the collected numbers of appearances from the story (column `$N(E_i)$' of Table \ref{piglethaffalunmp}), the red graph (Series 2) is a Bose-Einstein distribution model for these numbers of appearances (column `B-E mod' of Table \ref{piglethaffalunmp}), and the green graph (Series 3)  is a Maxwell-Boltzmann distribution model (column `M-B mod' of Table \ref{piglethaffalunmp}).}
\label{piglethaffalunmpgraphfigure}
\end{figure}
For the Winnie the Pooh story we find $E = 242{,}891$. Columns one, two, three and six contain the experimental data we retrieved from the Winnie the Pooh story. Some of these data are completely independent of the radiation scheme we propose for human cognition by analogy with the radiation of light, such as for example the numbers of words that are the same and the total number of words of the story. Other experimental data are not totally independent of this radiation scheme. The way we choose the energy levels, and how we add words to well-defined energy levels influences the values of these energies in an obvious way. Let us note, however, that it is difficult to imagine a more simple radiation scheme, but we will return to this later, for a real surprise awaited us in this respect.

\begin{figure}[h!]
\begin{center}
\includegraphics[width=10cm]{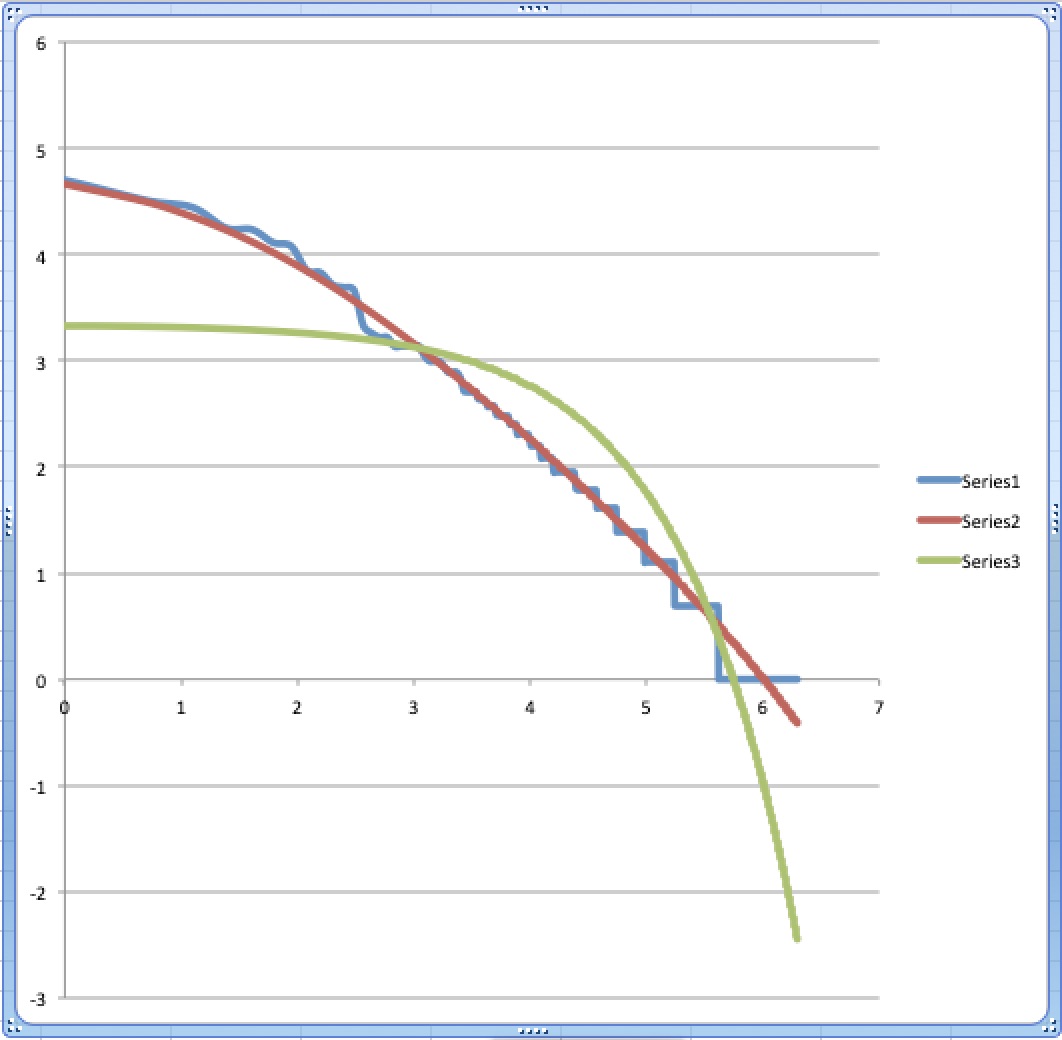}
\end{center}
\caption{Representation of the $\log / \log$ graphs of the `numbers of appearances' and their Bose-Einstein and Maxwell-Boltzmann models. The red and blue graphs coincide almost completely in both whereas the green graph does not coincide at all with the blue graph of the data. This shows that the Bose-Einstein distribution is a good model for the numbers of appearances, while the Maxwell-Boltzmann distribution is not.}
\label{piglethaffalunmploggraphfigure}
\end{figure}

\begin{figure}[h!]
\begin{center}
\includegraphics[width=10cm]{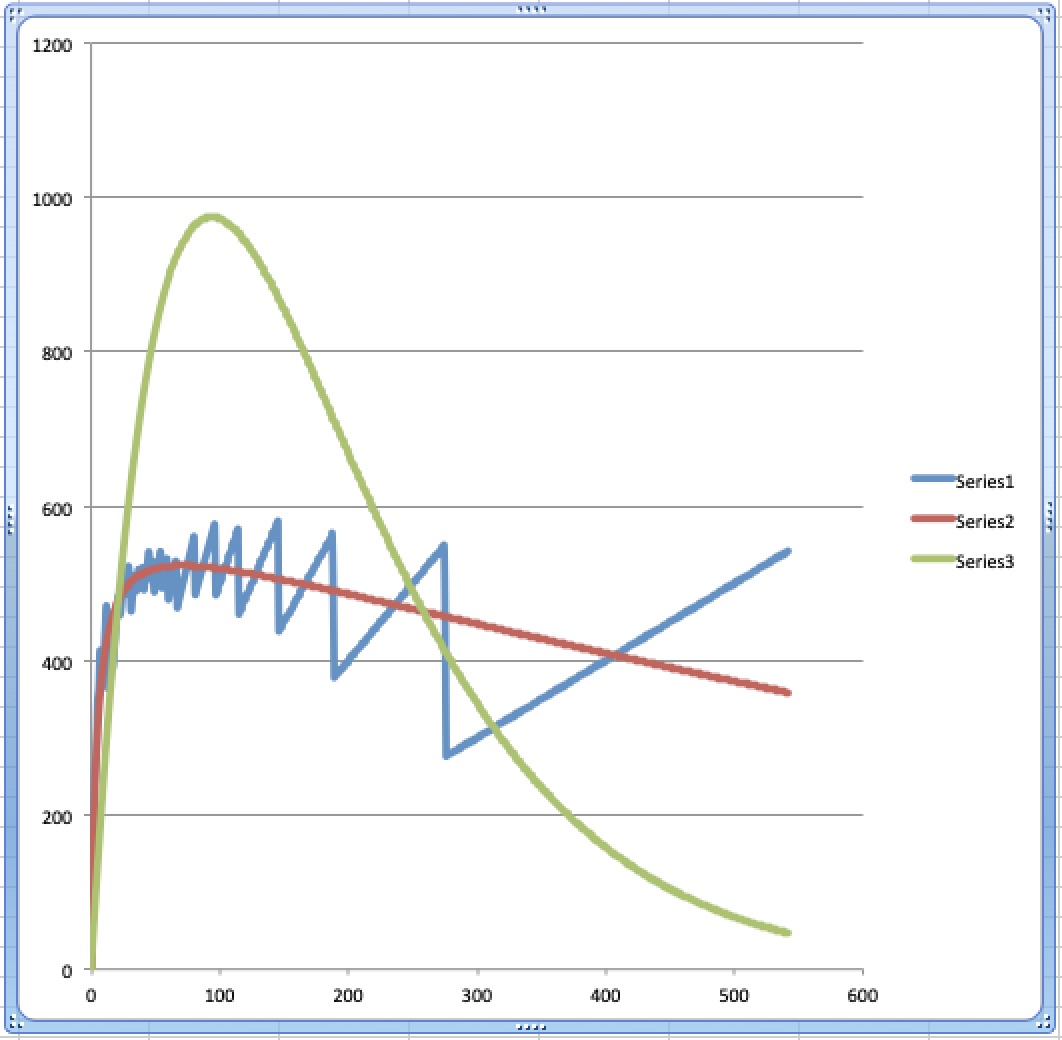}
\end{center}
\caption{A representation of the `energy distribution' of the Winnie the Pooh story `In Which Piglet Meets a Haffalump' \citep{milne1926} as listed in Table \ref{piglethaffalunmp}. The blue graph (Series 1) represents the energy radiated by the story per energy level (column `$E(E_i)$' of Table \ref{piglethaffalunmp}), the red graph (Series~2) represents the energy radiated by the Bose-Einstein model of the story per energy level (column `Energ B-E' of Table \ref{piglethaffalunmp}), and the green graph (Series 3) represents the energy radiated by the Maxwell-Boltzmann model of the story per energy level (column `Energ M-B' of Table \ref{piglethaffalunmp}).}
\label{piglethaffalunmpenergygraphfigure}
\end{figure}

Let us now consider columns four, five, seven and eight. They contain the results we obtained by testing both the Bose-Einstein distribution function (\ref{boseeinsteindistribution}) and the Maxwell-Boltzmann distribution function (\ref{maxwellboltzmanndistribution}) for their ability to model the experimental data contained in columns one, two, three and six within the proposed radiation scheme. And here we were rewarded with a big surprise, because when we tried the two functions, it turned out that the Bose-Einstein function showed a practically perfect modeling, after choosing the following values for the constants $A$ and $B$, $A = 1.0078$ and $B = 593.51$. Equally important for the possibility that our surprise could unfold into the feeling of being on the trail of a new important discovery, was that when we tried the Maxwell-Boltzmann function on the same experimental data from the Winnie the Pooh story, it did not lead to a good model at all, on the contrary, the Maxwell-Boltzmann function deviated very much from the experimental data. In addition to the data in Table \ref{piglethaffalunmp} itself, this result is very convincingly illustrated by the graphs we attached with the data contained in Table \ref{piglethaffalunmp}. In Figure \ref{piglethaffalunmpgraphfigure}, using a blue, red, and green graph, respectively, the experimental data (column three of Table \ref{piglethaffalunmp}), the Bose-Einstein function as a model (column four of Table \ref{piglethaffalunmp}), and the Maxwell-Boltzmann function as a model, are depicted.  Although in this Figure \ref{piglethaffalunmpgraphfigure} we can already clearly see how the Bose-Einstein function gives a very good description of the experimental data, and certainly in comparison to the really poor modeling of the Maxwell-Boltzmann function, this becomes even more clear when we calculate a log/log graph of the values in Figure \ref{piglethaffalunmpgraphfigure}. Such a log/log graph of the three graphs in Figure \ref{piglethaffalunmpgraphfigure} can be seen in Figure \ref{piglethaffalunmploggraphfigure}, where thus from both the $x$-axis and the $y$-axis simply the natural logarithms are calculated of the values in Figure \ref{piglethaffalunmpgraphfigure}. Even more manifestly than this was already the case in Figure \ref{piglethaffalunmpgraphfigure}, Figure \ref{piglethaffalunmploggraphfigure} illustrates how the Bose-Einstein function produces a practically perfect model and how the Maxwell-Boltzmann function totally fails to do so. 
In Figure \ref{piglethaffalunmpenergygraphfigure} is a graph where the energy values, i.e. columns six, seven and eight, are represented, again blue for the experimental data, red for the Bose-Einstein model and green for the Maxwell-Boltzmann model. Although, since multiplications take place, here the discreteness of the experimental data, while both the Bose-Einstein model and the Maxwell-Boltzmann model are continuous, is more noticeable, it is nevertheless clear here too that the Bose-Einstein model represents the correct evolution of the energetic radiation, while the Maxwell-Boltzmann model does not do so at all. At this point in the description of our result, it is important to emphasize again that we are not talking about an `approximation' here. The two experimental data for the Winnie the Pooh story, namely that the total number of words is equal to $N=2655$ and that the total energy radiated by the story is equal to $E=242{.}891$, unambiguously determine the constants $A$ and $B$, or $C$ and $D$, such that for both, Bose-Einstein, and Maxwell-Boltzmann, there is only one unique solution that satisfies these experimental constraints, the size of $N$, and the size of $E$. 
That the Bose-Einstein function, as defined in (\ref{boseeinsteindistribution}) (the red graph in Figure \ref{piglethaffalunmpgraphfigure}, Figure \ref{piglethaffalunmploggraphfigure}, and Figure \ref{piglethaffalunmpenergygraphfigure}), coincides with the experimental function (the blue graph in Figure \ref{piglethaffalunmpgraphfigure}, Figure \ref{piglethaffalunmploggraphfigure}, and Figure \ref{piglethaffalunmpenergygraphfigure}), is therefore a very special event, and not at all a matter of `approximation' or `fitting of parameters'. Especially when this is not the case at all for the Maxwell-Boltzmann function, as defined in (\ref{maxwellboltzmanndistribution}) (the green graph in Figure \ref{piglethaffalunmpgraphfigure}, Figure \ref{piglethaffalunmploggraphfigure} and Figure \ref{piglethaffalunmpenergygraphfigure}). 

Admittedly, it was also possible that the agreement of the experimental data of the Winnie the Pooh story with the Bose-Einstein statistics was a lucky coincidence, and so we immediately started looking for other stories to see if this phenomenon would be repeated. Because we wanted to avoid having to manually classify words from other stories into our radiation scheme again, we planned in parallel to develop a program that would perform this ordering with the computer. 
And then a second surprise happened with potentially far-reaching consequences, we found after some searching around on the World-Wide Web that such a program already existed, and more, that it was being used by researchers interested in Zipf's law in human language \citep{zipf1935,zipf1949}.  But more importantly, we discovered that the scheme proposed by Zipf was very similar, and in the simplest case, even completely the same than our scheme. It is true that the notions of `energies' and the idea of `radiation' were not introduced, but the mathematical form of the two schemes was the same. Zipf's law is an empirical law, and although theoretical foundations are tried here and there, there is certainly no theoretical derivation that is accepted by consensus as a theoretical foundation for Zipf's law. If we consider the log/log graph that goes along with the Winnie the Pooh story, i.e., the graph shown in Figure \ref{piglethaffalunmploggraphfigure}, we do indeed see a curve that does not deviate much from a straight line, and this is what was particularly noted by Zipf. 
So when we began to examine texts from other stories, there was the additional question of whether there would be a connection there as well, and whether the log/log graphs that we will look at are those that also interested Zipf from a purely empirical standpoint. For the many subsequent texts we examined, of many different stories, shorter as well as longer than the Winnie the Pooh story, and also of the length of novels, again and again we found an equally strong confirmation of the presence of a Bose-Einstein statistics \citep{aertsbeltran2020}. Zipf's law \citep{zipf1935,zipf1949}, and its continuous version, Pareto's law \citep{pareto1897}, both turn up in all kinds of situations of human-created systems, the rankings of cities according to their size \citep{gabaix1999}, the rankings of incomes from low to high \citep{aokimakoto2017}, of wealth, of corporation sizes, but also in rankings of number of people watching the same TV channel \citep{erikssonetal2013}, cells `transcriptomes' \citep{lazardietal2021}, the rankings of notes in music \citep{zanette2006} and so on ... so certainly it is not only in human language that this law emerges. That Zipf's and Pareto's laws appear so prominently in many areas of human presence is a hint that our theoretical underpinning of these laws by means of a radiation and quantization scheme may also find wide application in these other areas, and not only concerns the structure of human language. The connection we make in the next Section \ref{meaningdynamics} with a `meaning dynamics' also already points in the direction of a possible wide application in these other areas.

\section{A dynamics of meaning}
\label{dynamicsofmeaning}

We wish in this Section to arrive at some new insights that account for why we dared to present the comparison between electromagnetic radiation and how it interacts with matter, and human language and how it interacts with the human mind, i.e., the dynamics of human cognition, as more than just a superficial metaphor. In \citet{aertsbeltran2020}, the first paper where we wrote down our findings, we worked out a comparison between the human language and a gas composed of bosonic atoms close to absolute zero, or, more precisely, in the temperature region where the Bose-Einstein condensate phenomenon begins to manifest itself. The reason we then opted for that comparison, while primarily Planck's radiative law, as we outline it here in Section \ref{radiationlawhumanlanguage}, had been our inspiration, was, because we thought in that way we could more easily gain insight into `what those energies of words really represent'. After all, photons always move at the speed of light, and have mass equal to zero, whereas atoms are closer to our near material reality. Probably we were also influenced by the spectacular aspects of the Bose-Einstein condensation, realized in 1995, and by now well mastered by experimental quantum physicists in many labs around the world \citep{cornellwieman2002,ketterle2002}. We should also note here that our activity within the research area of quantum cognition has always been two-fold, consisting of at least two strands. On the one hand, the use of the quantum formalism in human cognition leads to the possibility of describing kinds of dynamics within human cognition, e.g. decisions, but often also reasonings that are called `heuristics', or `fallacies', not satisfying ordinary logic- and rationality-based dynamics. Also, in this strand within quantum cognition we have always tried to `understand' what exactly is going on, `heuristics' and even `fallacies' may be `logics and rationalities' of a hidden deeper nature of human cognition. The second strand, and we certainly do not consider this less important, is to better understand quantum mechanics as applied to the description of the physical micro-world, by analyzing the application of quantum mechanics in human cognition and language. And certainly in our investigation of the presence of Bose-Einstein statistics in human cognition, this second strand is very present. So the idea is that, `if we can understand how Bose-Einstein forms within human cognition and language, then perhaps that can give us insights into what Bose-Einstein means for the physical micro-world'. During the work and research in this comparison of human language with a boson gas in a temperature close to absolute zero, we became fascinated by the question `how it was possible that Albert Einstein' wrote a paper describing the Bose-Einstein condensate as early as 1924, i.e. before the emergence of modern quantum mechanics in 1925-1926. As we examined this question more closely, a world of thought, discourse, correspondence opened up, which step by step, by reading more and more about it, we began to find scientifically relevant to our finding of the presence of Bose-Einstein statistics in human cognition and language. 

The episode where the Indian physicist Satyendra Nath Bose wrote a letter to Einstein announcing a new proof for Planck's law, and how Einstein was so excited by Bose's new calculation that he translated Bose's article into German, and arranged for it to appear in the then leading journal `Zeitschrift f\"ur Physik' \citep{bose1924} is well known. However, what really took place, why for example Einstein was so enthusiastic, and why he applied Bose's method to an ideal gas right way, and wrote three articles about it \citep{einstein1924,einstein1925a,einstein1925b}, where in the second one he predicted the existence of a Bose-Einstein condensate, is already much less known. 
Historical sources, which have since been studied in great detail by historians of science and philosophers of science, including letters of correspondence between the scientists involved, show that three scientists played the leading role reflecting on the Planck's law of radiation. Max Planck, Albert Einstein and Paul Ehrenfest, with a one-time but influential appearance by Satyendra Nath Bose, dominated the discussions about the nature of electromagnetic radiation between 1905 and 1925, the period often called the Old Quantum Theory  \citep{howard1990,darrigol1991,monaldi2009,perezsauer2010,gorroochurn2018}. We want to draw on the interesting events, reflections, and discussions that took place during that period to point out the connections not only to our own research, but also to the endeavor of what we have now come to call `Quantum Cognition' as a domain of research.

First of all, let us note that Max Planck was already inspired by Boltzmann's statistical view of thermodynamics in the original formulation of his radiation law \citep{planck1900}. Indeed, he proposed a model for the radiation of a black body that used a theory originally developed for the description of gases by Boltzmann. The latter had already used his own thermodynamics to study light radiation \citep{boltzmann1978}, and in this sense had also influenced Wien \citep{wien1897}, on which Planck built further. It is sometimes claimed that Planck's law brought a solution to the ultraviolet catastrophe, namely the prediction made by a law derived by Lord Rayleigh, that light in higher frequencies would radiate more and more, which clearly did not agree with the experiments. However, historical sources indicate that this is not how things happened, Rayleigh's law was only published in June 1900, when Planck had already fully enrolled, in the footsteps of Boltzmann and Wien, to work out a thermodynamic description for the radiation of a black body \citep{klein1961}. We already noted in Section \ref{radiationlawhumanlanguage}, that Planck did not think of quantization, and that it was Einstein in 1905 who brought this notion to the stage. However, the difference in interpretation of what was going on between Planck and Einstein, with also Ehrenfest engaging in reflecting on these differences, was much deeper than that. It is worth noting that Ehrenfest was the principal student of Ludwig Boltzmann in Vienna and was also a close friend of Einstein. With Einstein's support he had succeeded Hendrik Lorentz at the University of Leiden in the Netherlands and Einstein would visit Leiden regularly to discuss with his friend. Let us make clear some scientific aspects of the situation under consideration before we pick out what is scientifically important to us from these discussions. What Planck, Einstein and Ehrenfest were reflecting on, following Boltzmann, starts from a formula with combinatorial quantities in it, derived from a combinatorial reasoning, often formulated from an imagining of situations of `particles distributed among baskets' -- think of the reasoning of the farmer on the farm who picks out two cats and two dogs by putting them in two baskets, as we described in Section \ref{boseeinstein}. To then approximate this combinatorial formula by a continuous function, limits are applied that are very good approximations for when many particles are involved. And indeed many particles are always involved when statistical mechanics is used for thermodynamic purposes. Just as the three scientists were in complete agreement that the radiation law had to be derived using statistical thermodynamics, there was likewise no difference of opinion regarding any aspect of this limit calculation, which by the way is standard, and can be found in any book on statistical thermodynamics \citep{huang1987}. So then what were the discussions about, and what did the three disagree about? 

We can best illustrate the disagreement by going back to the example of the two entities that can be in two states, as we considered in Section \ref{boseeinstein}. And to stay close to the subject of our paper, we take for the two entities, two concepts {\it Animal}, and the two states in which each concept {\it Animal} can be found are, {\it Cat} and {\it Dog}. Planck imagined entities he called `resonators', as elementary constituents of the electromagnetic radiation he wanted to model. The energy elements were then `distributed among the various resonators' in Planck's representation of the situation. And, since energy elements have no identity, Planck used a combinatorial formula, which is this one of `identical particles distributed among baskets', before making the limit transition to the continuous function. That is the reason why, after the limit procedure, Planck finds a continuous function that possesses the Bose-Einstein form, such as (\ref{boseeinsteindistribution}), which is indeed to be found back in (\ref{planckradiation}), as we already remarked in Section \ref{radiationlawhumanlanguage}. 
Einstein, however, already in his 1905 article \citep{einstein1905} on the photo electric effect, was insistent on considering the energy elements as quanta, which he even wished to grant an existence of their own in space, and Ehrenfest supported him in this way of interpreting Planck's radiation law. Over the years there was also a growing understanding that when considering the light quanta as entities in themselves, like atoms and molecules of a gas, Planck's derivation of the radiation formula exposed a fundamental problem. The probabilities for the relevant micro states, which were necessary for Boltzmann's theory in the calculation of entropy, did not appear to be statistically independent for the different light quanta. More to the point, if the probabilities for the micro states for different light quanta were assumed to be independent, the radiation law again became that of Wien, and in 1911 Ehrenfest could derive this result formally \citep{ehrenfest1911,monaldi2009}. In the years that followed it became more and more clear what the nature of the disagreement was between Planck on the one hand and Einstein and Ehrenfest on the other, when the lack of statistical independence became the focus of this disagreement. We have already mentioned what is meant by this lack of statistical independence using the example of the {\it Two Animals}, which can be {\it Cat} or {\it Dog}, in Section \ref{boseeinstein}. The Maxwell-Boltzmann probabilities of this situation are $1/4$ for two cats, $1/4$ for two dogs, and $1/2$ for a cat and a dog, while the Bose-Einstein probabilities are given by $1/3$ for two cats, $1/3$ for two dogs, and $1/3$ for a cat and a dog. It is easily to understand that the latter cannot result from an independent choice between {\it Cat} and {\it Dog}. For Einstein, who wanted to consider the energy packets in Planck's formula as independently existing light quanta, this lack of independence was a major problem. Meanwhile the notion of light quanta was gaining ground among other scientists. For some, the idea that light particles formed in conglomerates of light molecules began to make sense. However, there were also numerous physicists who followed the view of Planck, and explained the peculiar statistics of the energy elements required by Planck's law of radiation as being rather useful `fictions' of the `resonator model', and awaiting further discoveries related to it \citep{howard1990,perezsauer2010}.

Things took a new turn when Einstein received a letter from Satyendra Nath Bose, who claimed to have found a new derivation of Planck's radiation law. Bose had tried to publish his paper but had failed, and he sent it to Einstein, suspecting that the latter would be interested, Bose was indeed aware of the discussions about Planck's radiation law that were going on. Einstein was intrigued by Bose's calculation and translated Bose's article into German sending it to Zeitschrift f\"ur Physik, where it was published \citep{bose1924}. However, more, immediately he set to work introducing Bose's approach for a model of an ideal gas, about which he published an article the same year, and two additional articles in January of the following year \citep{einstein1924,einstein1925a,einstein1925b}. Ehrenfest was not at all enthusiastic about Bose's method and made this clear to a fellow physicist, writing the following in a letter dated October 9, 1924, to Abram Joff\'e: ``Precisely now Einstein is with us. We coincide fully with him that Bose's disgusting work by no means can be understood in the sense
that Planck's radiation law agrees with light atoms moving independently (if they move independently one of each other, the entropy of radiation would depend on the volume not as in Planck, but as in W. Wien, i.e. in the following way: $\kappa \log V^{E \over h}$).'' (in \citet{moskovchenkofrenkel1990}, pp. 171--172). Einstein was silent on the problem of the statistical dependence of the light quanta leading to Planck's law of radiation using Bose's method in his first quantum gas article \citep{einstein1924}. The second article \citep{einstein1925a} he wrote during a stay in Leiden with Ehrenfest, and in it he openly states, ``Ehrenfest and other colleagues have objected, regarding Bose's theory of radiation and my analogous theory of the ideal gas, that in these theories the quanta or molecules are not treated as statistically independent entities, without this circumstance being specifically pointed out in our articles. This is entirely correct." (in \citet{einstein1925a} page 5).  

It is important to note that, parallel to these events, Louis de Broglie had come forward with the hypothesis of `a wave character also for material particles' \citep{debroglie1924}, and, although Einstein only learned of de Broglie's writings after his contacts with Bose, he had read de Broglie's thesis with great interest before writing his second article on the quantum gas~\citep{einstein1925a}. In this article, it becomes clear that he sees the statistical dependence of gas atoms as caused by a `mutual influence', the nature of which is as yet totally unknown. However, he also mentions that the wave-particle duality, brought out in de Broglie's work, might play a role in unraveling it \citep{einstein1925a,monaldi2009}. Not much later, in a letter to Erwin Schr\"odinger, who had begun to take an interest in the quantum gas at this time \citep{schrodinger1926a}, Einstein wrote the following passage: ``In Bose's statistics, which I have used, the quanta or molecules are not treated as independent from one another. [ \ldots ] According to this procedure, the molecules do not seem to be localized independently from each other, but they have a preference to be in the same cell with other molecules. [\ldots ] According to Bose, the molecules crowd together relatively more often than under the
hypothesis of statistical independence.'' (in~\citet{monaldi2009} page~392, see also Figure 1 in \citet{monaldi2009}, where this passage is shown in Einstein's handwriting). So here Einstein explained in a crystal clear way the nature of this statistical dependence to Schr\"odinger. Several letters between Einstein and Schr\"odinger in the late year of 1925 about Einstein's quantum gas, and in parallel Schr\"odinger taking note of de Broglie's work on the quantum theory of gases \citep{debroglie1924}, were instrumental in Schr\"odinger's formulation of wave mechanics \citep{schrodinger1926b,howard1990}. When Einstein realized what it meant that Schr\"odinger's matter waves were defined on the configuration spaces of the quanta, which made these waves fundamentally different from de Broglie's waves, it marked the turning point for him, where he would step away from the new more and more abstract mathematically defined quantum formalism \citep{howard1990}.  

In modern quantum mechanics, the Bose-Einstein statistics is directly connected to the `exchange symmetry' applied to the quantum states, i.e. the unit vectors of the complex Hilbert space. In this sense, `identity' and `indistinguishability' seem to be the unique new way of understanding the Bose-Einstein statistics, but is that really the case? Is a blue photon indistinguishable from a red photon even if we can distinguish the two with our human eyes? A blue photon and a red photon possess different amounts of energy, and in a radiation scheme such as we have proposed here, both will therefore be classified in a different energy level. By the way, it is always good when it comes to notions that are difficult to grasp, to find out how quantum experimentalists deal with them. Especially since there is now interest in using the `indistinguishability' of photons with the intention of fabricating `entangled photons' that could be used as qubits for quantum computing. It then becomes clear, just by checking how quantum experimentalists deal with it, that blue photons are indeed considered `distinguishable' from `red photons'. We describe in \citet{aertsbeltran2020} how indistinguishable photons are prepared and treated by quantum experimentalists, but in short, how blue photons are distinguishable from red photons is quite the same as we have also classified different words in different energy levels within the radiation scheme we have proposed for human cognition and language. And indeed, different words are not indistinguishable, even by definition of what they are. Yet it is precisely there, with such not indistinguishable photons and words that the Bose-Einstein statistics intervenes by thoroughly attaching other probabilities of occurrence to them. And not just `other probabilities of occurrence', but `probabilities of occurrence that cannot come from independent choices for the individual cases'. Consider again the example of the choice between a cat and a dog, and how according to the Maxwell-Boltzmann statistics this gives rise to $1/2 \cdot 1/2 = 1/4$ for `two cats', $1/2 \cdot 1/2 = 1/4$ for `two dogs', and $1/2 \cdot 1/2 + 1/2 \cdot 1/2 = 1/2$ for `a cat and a dog', probabilities of occurrence which `do' obtain as products of the probabilities of occurrence of the individual cases, choosing with probability $1/2$ for `one cat' and probability $1/2$ for `one dog'. And how the Bose-Einstein statistics of this situation, with probabilities of occurrence $1/3$, for `two cats', $1/3$ for `two dogs', and $1/3$ for `a dog and a cat', do not allow such a decomposition into products of individual probabilities of occurrence at all. It is this lack of independence that Einstein found so annoying, and to which he suddenly, as if he had grown tired of resisting it, admitted in the method of calculation that Bose presented to him. It is argued in \citet{howard1990}, with numerous excerpts from articles, lectures, and correspondences of Einstein that this phase nevertheless also marked his turnaround from quantum mechanics. Einstein saw no other possibility than to concede that the probabilities of occurrence of different quanta of light were not independent, because the experiments failed for a Maxwell-Boltzmann version of Planck's radiation law, which was, after all, Wien's radiation law. But Einstein believed that there was a specific force at work that caused these peculiar probabilities of the Maxwell-Boltzmann statistics to come into existence. When modern quantum mechanics came into full development in the following years, and Erwin Schr\"odinger formulated his wave mechanics on the basis of waves in the configuration space of particles, causing the phenomenon we now call `entanglement', Einstein found this a bridge too far, and began the resistance that would lead to the famous article with the EPR paradox \citep{einsteinpodolskyrosen1935}.

In the studies in which it was our intention to identify the Bose-Einstein statistics in human cognition and language, and which we described in Section \ref{boseeinstein}, we proposed that `a cat and a dog' would indeed be chosen no more frequently than `two cats' or `two dogs'. And we insinuated that this would be due to the `context in which the choice takes place' to be such that the alternatives from which to choose are restricted to the three alternatives, `two cats', `two dogs', or `a cat and a dog', thus resulting in 1/3, 1/3, and 1/3 as probabilities of choice. Whereas, if one forms the context in which the choice takes place with the four alternatives `two cats', `two dogs', `one cat and one dog', and `one dog and one cat', then one gets the natural context for a Maxwell-Boltzmann type statistics, where then the probabilities for the choices will be 1/4, 1/4 and 1/2. But does this provide a complete and satisfying explanation for the presence of the Bose-Einstein statistics? Let's revisit our analysis of the Winnie the Pooh story `In Which Piglet Meets a Haffalump' to show that there may be something else going on after all. We consider the words in Table \ref{piglethaffalunmp}. There we see that the word {\it Piglet} is classified with energy level $8$ and occurs $47$ times in the story. The word {\it First} is classified with energy level $71$ and occurs $7$ times in the story, while the word {\it Year} is classified with energy level $541$ and occurs $1$ time in the story. Now suppose we could ask Alan Alexander Milne to write a few extra paragraphs specifically for this Winnie the Pooh story. It is easy to understand that in these new paragraphs the probability of the word {\it Piglet} appearing is much greater than the probability of the word {\it First} appearing, and the probability for that is much greater again than the probability of the word {\it Year} appearing. Why is this so? The reason is that the story carries `meaning', and the words that have more affinity with this meaning that the story carries have a greater chance of appearing when the story is written. But, another way of expressing the same is to state that `meaning' is a force that makes the same words attract each other to clump together as a result of this `meaning force'. By identifying `meaning' as the cause for `the clumping of the same words in the text of a story', have we discovered the force, this one that Einstein called a `mysterious unprecedented force', that causes photons to clump into the same states within light?

We will now consider an example of a story consisting of a small number of words such that we can use the original exact combinatorial formulas to describe and analyze it. Thus, our analysis this time will be `exact' without being obliged, due to the large number of words, to proceed to the use of the continuous functions, such as (\ref{boseeinsteindistribution}) and (\ref{maxwellboltzmanndistribution}), which are approximations to these exact original combinatorial formulas, to identify the difference between Bose-Einstein and Maxwell-Boltzmann. This will allow us to gain more insight yet into `what exactly is going on'. We will see that we can likewise better identify the subtle structure of this `meaning dynamics'. 
It will also allow us to give the `Fermi-Dirac statistics' a place within our global analysis. We have proposed the situation we wish to analyze in Table \ref{fourwordssevenenergylevels}. We consider `four words' distributed over `seven energy levels' in such a way that for each configuration `the total energy of this configuration is equal to seven'. Let us first describe what we find in the first part of Table \ref{fourwordssevenenergylevels}, under the heading `Four Words Distributed over Seven Energy Levels'. The first column shows the numbering of the energy levels, including an energy level $0$, and then $1$, $2$, $3$, $4$, $5$, $6$ through $7$. Column two shows a first possible configuration of four words with total energy equal to seven, it is the configuration where one word possesses energy level $7$, and three words possess energy level $0$. Column three shows a second possible configuration, this time one word in energy level $6$, one word in energy level $1$, and two words in energy level $0$. Indeed, this again gives four words distributed among the seven energy levels, and such that the total energy is equal to seven. The following columns show the other possible such configurations, always four words distributed over seven energy levels such that the total energy is equal to seven. Each such configuration is called a macro state, it will become clear in the following why we introduce this terminology. Table \ref{fourwordssevenenergylevels} shows that we can construct eleven such different configurations in this way, so there are eleven such macro states for four words distributed over seven energy levels with total energy equal to seven. In the third row of the first part of Table \ref{fourwordssevenenergylevels}, we have numbered these macro states, from 1 to 11. 
Let us now explain what the numbers represent that are found in Table \ref{fourwordssevenenergylevels} on the second row of the first part of the table, which we have called `Maxwell-Boltzmann numbers', written `M-B numbers' on that second row of the table. 
The first of those numbers, in the second column of the table, i.e. belonging to the first macro state, is equal to $4$. It is the number of ways that the configuration of this first macro state can form if an underlying interchange of the four words in play leads to a distinguishable state, without counting interchanges of words that are in the same energy level, as they are by definition indistinguishable. 
And, if we thoroughly let sink in what is being formulated here, we understand that there are indeed $4$ such micro states. Namely, any of the three words that are in energy level $0$ can be moved to energy level $1$, to lead to a state that is basically distinguishable from the previous one. That gives three new distinguishable states plus the one we departed from before we started moving, thus giving a total of $4$ basically distinguishable states. We already told about the combinatorial formulas we would use, and, to arrive at this number $4$, there is such a formula. It is given by dividing $4!$ -- the number of possible differences by exchanging all four words -- dividing by $3!$ -- expressing that the exchanges of the indistinguishable words in the same energy level are not counted. The general combinatorial formula that gives us for each macro state the number of micro states that exist underlying it, and that are in principle distinguishable from each other, is the following
\begin{eqnarray} \label{microstates}
N_{M-B} = {N! \over N_0! \cdot N_1! \ldots N_i! \ldots N_7!}
\end{eqnarray}
where $N$ is the number of considered words, hence $N=4$ in our case, and $N_i$ is the number of words in energie level $i$. Remember that $0! = 1$, which is important for the formula to apply correctly to all macro states, even when 0 words are present in a specific energy level. We calculate from (\ref{microstates}) the numbers of micro states for each macro state, in Table \ref{fourwordssevenenergylevels} to be found in the second row denoted by `M-B numbers'. Let's take a moment to consider the largest number of micro states we obtain in this way, namely the $24$ belonging to macro state number six. Indeed, we wish to emphasize again why we consider these micro states as `in principle' distinguishable. 
For macro state number six, each of the four words belong to a different energy level and are therefore, by definition, different words. Let us give an example of such a piece of text consisting of four different words. `Dog loves white cat' is such a sentence. Now if we interchange `dog' and `cat', this sentence becomes `Cat loves white dog', and this is indeed a different sentence, which we have no problem distinguishing from the previous one. This example makes clear how this `distinguishability' is akin to what we were thinking about with our simple example in Section \ref{boseeinstein}, of choosing two animals, with the option of choosing a cat or choosing a dog. We can probably find examples of sentences with four words that we do not treat as distinguishable within a given context, and that is why we added the `in principle'. The total number of micro states is given by the sum of all micro states per macro state, and for this case of four words distributed over seven energy levels, such that the total energy is always equal to seven, this total number of micro states is equal to $120$. We can now move on to the second part of Table \ref{fourwordssevenenergylevels}, under the title `Maxwell-Boltzmann'.

\begin{table}
\centering
\caption{The data of an example consisting of four words distributed over seven energy levels with total energy equal to seven. The table consists of four parts.The first part gives the raw data. The other three parts give the calculated data to arrive at the Maxwell-Boltzmann, Bose-Einstein and Fermi-Dirac function respectively.}
\label{fourwordssevenenergylevels}
\begin{tabular}{p{2.4cm}p{0.9cm}p{0.9cm}p{0.9cm}p{0.9cm}p{0.9cm}p{0.9cm}p{0.9cm}p{0.9cm}p{0.9cm}p{0.9cm}p{0.9cm}}
\hline
\hline
\multicolumn{12}{c}{Four Words Distributed over Seven Energy Levels} \\
\hline
\hline
M-B numbers& 4 		& 12 	& 12  	& 12		& 12 	& 24 	& 4 		& 12 	& 12 	& 12 	&	4 \\
\hline
Macro states	& 1 	& 2   	& 3    & 4 		& 5 		& 6 		& 7  		& 8 		& 9 	& 10 & 11  \\
\hline
0	& 3 		& 2   	& 2    & 1 	& 2 		& 1 		& 0		&1 		& 1 		& 0 		& 0 \\
1	& 0 		& 1   	& 0    	& 2		&0 		& 1 		& 3		&1 		& 0 	& 2 	& 1 \\
2	& 0 	& 0   	& 1    	& 0 		& 0 		& 1 		& 0		&0 		& 2 		& 1 	& 3 \\
3	& 0 	& 0   	& 0    	& 0	 	& 1 		& 0 		& 0		&2 		& 1 		& 1 	& 0 \\
4	& 0 		& 0   	& 0    	& 0 	& 1 		& 1 		& 1		&0 		& 0 	& 0 		& 0 \\
5	& 0 		& 0   	& 1    	& 1 		& 0 		& 0 		& 0		&0 		& 0 	& 0 	& 0 \\
6	& 0 		& 1  		& 0    	& 0 		& 0  	& 0  		& 0	&0  		& 0  		& 0 	& 0 \\
7	& 1		& 0  		& 0    	& 0 		& 0  	& 0  		& 0	&0  		& 0  	& 0 		& 0 \\        
\hline
\hline
\multicolumn{12}{c}{Maxwell-Boltzmann} \\
\hline
\hline
M-B weights &	0.03	& 0.1  	&	0.1   & 0.1   	& 0.1  	& 0.2  	& 0.03  & 0.1  	&	0.1   & 0.1 	& 0.03 \\
\hline
Macro states	& 1		& 2   	& 3    	& 4 		& 5 		& 6 		& 7  		& 8 		& 9 		& 10 	& 11  \\
\hline
0					& 0.1 	& 0.2   	& 0.2   & 0.1		& 0.2 	& 0.2 	& 0		& 0.1 	& 0.1 	& 0 		& 0 \\
1					& 0 		& 0.1   	& 0    	& 0.2	&0 		& 0.2 	& 0.1	& 0.1 	& 0 		& 0.2 	& 0.03 \\
2					& 0 		& 0   	& 0.1    & 0 		& 0 		& 0.2 	& 0		& 0 		& 0.2 	& 0.1 	& 0.1 \\
3					& 0 		& 0  		& 0    	& 0 		& 0.1 	& 0 		& 0		& 0.2 	& 0.1 	& 0.1 	& 0\\
4					& 0 		& 0   	& 0    	& 0 		& 0.1 	& 0.2 	& 0.03	&	0 		& 0 		& 0 		& 0 \\
5					& 0 		& 0   	& 0.1   	& 0.1 	& 0 		& 0 		& 0		&	0 		& 0 		& 0 		& 0 \\
6					& 0 		& 0.1  	& 0    	& 0 		& 0  		& 0  		& 0		&0  		& 0  		& 0 		& 0 \\
7					& 0.03	& 0  		& 0    	& 0 		& 0  		& 0  		& 0		&0  		& 0  		& 0 		& 0 \\
\hline
\hline
\multicolumn{12}{c}{Bose-Einstein} \\
\hline
\hline
B-E weights	&	0.09	&	0.09	&	0.09  &	0.09  &	0.09 	&	0.09	&	0.09	&	0.09 	&	0.09 &	0.09 &	0.09\\
\hline
Macro states	& 1		& 2   	& 3    	& 4 		& 5 		& 6 		& 7  		& 8 		& 9 		& 10 	& 11  \\
\hline
0					& 0.27 	& 0.18  & 0.18  & 0.09	& 0.18 	& 0.09 	& 0		& 0.09 	& 0.09 	& 0 		& 0 \\
1					& 0 		& 0.09  & 0    	& 0.18	& 0 		& 0.09 	& 0.27	& 0.09 	& 0 		& 0.18 	& 0.09 \\
2					& 0 		& 0   	& 0.09  & 0 		& 0 		& 0.09 	& 0		& 0 		& 0.18 	& 0.09 	& 0.27 \\
3					& 0 		& 0  		& 0    	& 0 		& 0.09	& 0 		& 0		& 0.18 	& 0.09 	& 0.09 	& 0\\
4					& 0 		& 0   	& 0    	& 0 		& 0.09 	& 0.09 	& 0.09	& 0 		& 0 		& 0 		& 0 \\
5					& 0 		& 0   	& 0.09  & 0.09 	& 0 		& 0 		& 0		& 0 		& 0 		& 0 		& 0 \\
6					& 0 		& 0.09	& 0    	& 0 		& 0  		& 0  		& 0		& 0  		& 0  		& 0 		& 0 \\
7					& 0.09	& 0  		& 0    	& 0 		& 0  		& 0  		& 0		&0  		& 0  		& 0 		& 0 \\
\hline
\hline
\multicolumn{12}{c}{Fermi-Dirac} \\
\hline
\hline
F-D weights	&	0		&0.125&0.125	&0.125&0.125 	&	0.125&	0		&	0.125&	0.125&	0.125&	0\\
\hline
Macro states	& 1		& 2   	& 3    	& 4 		& 5 		& 6 		& 7  		& 8 		& 9 		& 10 	& 11  \\
\hline
0					& 0 		& 0.25  & 0.25 	&0.125& 0.25 	&	0.125& 0		&0.125	&0.125	& 0		& 0 \\
1					& 0 		&0.125	& 0    	& 0.25	& 0 		&0.125	& 0		&0.125& 0 		& 0.25 	& 0 \\
2					& 0 		& 0   	&0.125	& 0 		& 0 		&	0.125& 0		& 0 		& 0.25 	&	0.125& 0 \\
3					& 0 		& 0  		& 0    	& 0 		&0.125	& 0 		& 0		& 0.25 	&0.125	&	0.125& 0\\
4					& 0 		& 0   	& 0    	& 0 		&0.125 &	0.125& 0		& 0 		& 0 		& 0 		& 0 \\
5					& 0 		& 0   	&	0.125&0.125	& 0 		& 0 		& 0		& 0 		& 0 		& 0 		& 0 \\
6					& 0 		&	0.125& 0    	& 0 		& 0  		& 0  		& 0		& 0  		& 0  		& 0 		& 0 \\
7					& 0		& 0  		& 0    	& 0 		& 0  		& 0  		& 0		&0  		& 0  		& 0 		& 0 \\
\hline\\
\end{tabular}
\end{table}

\begin{figure}[h!]
\begin{center}
\includegraphics[width=10cm]{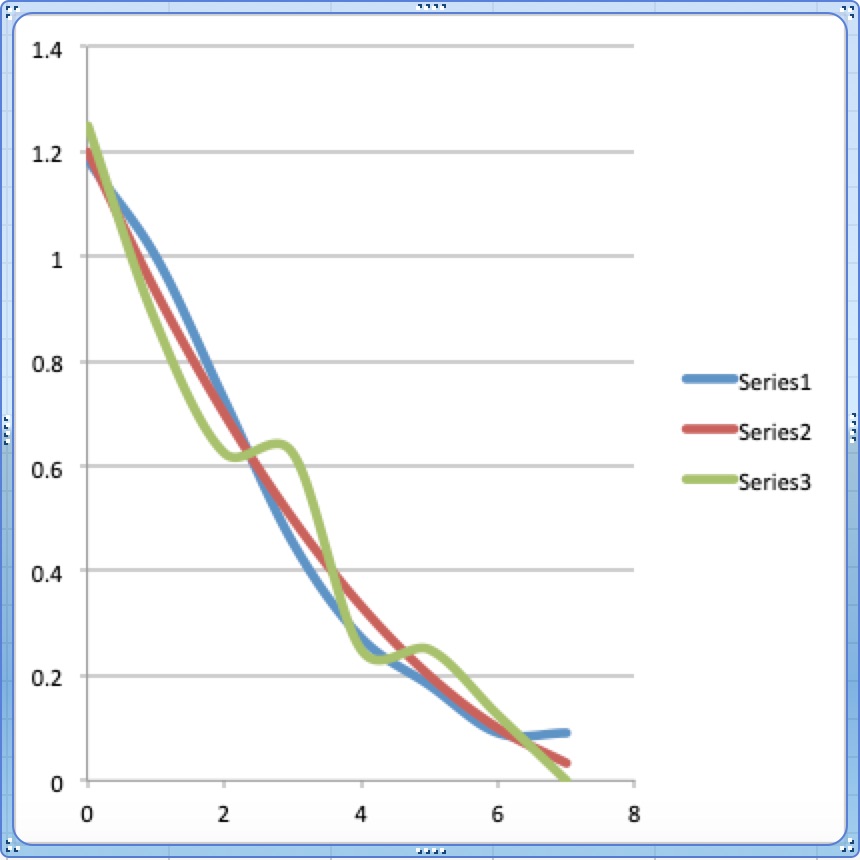}
\end{center}
\caption{
The blue, red, and green curves are the graphs of the Bose-Einstein, Maxwell-Boltzmann, and Fermi-Dirac functions for the situation of four words distributed over seven energy levels with the total energy equal to seven.}
\label{Bose-Boltzmann-FermiSevenFourEnergyLevelsWordsfigure}
\end{figure}

Here we make the calculations to arrive at the exact, i.e., only based on combinatorial formulae, version of what the continuous Maxwell Boltzmann function is, as given in (\ref{maxwellboltzmanndistribution}). 
In the first row of this second part of Table 2, we calculate the Maxwell-Boltzmann weights. To do this, we divide for each macro state, the number of micro states that this macro state contains, by the total number of micro states. That is a very natural way to assign a weight per macro state that probabilistically captures the number of micro states in that macro state, the sum of all the weights being equal to $1$. Or, to really use the image in which Ludwig Boltzmann looked at this situation, suppose all micro states are equally likely to be realized, then each Maxwell-Boltzmann weight of a macro state represents the probability that this macro state will be realized under a Maxwell Boltzmann statistics. The numbers in rows four through eleven of the second part of Table \ref{fourwordssevenenergylevels} give the weighted average values per energy level of the words present in each macro state. Let us look at one of them in detail, for example, the number located in the fifth row, the row of energy level $1$, and fifth column, the column of macro state $4$, which is equal to $0.2$. We obtained this number by multiplying the B-E weight of the macro state in this fifth column, i.e., macro state number $4$, and thus number $0.1$, by the number of words located in the first part of Table \ref{fourwordssevenenergylevels}, i.e., $2$ words, and hence this gives $0.2$. So this $0.2$ is the weighted average number of words contributed by the underlying micro states of this macro state to energy level $1$.  If we add up all these weighted averages obtained in the same way over all the fifth row, that is, the row of energy level $1$, we get the value of the Maxwell-Boltzmann function in energy level $1$. And if we do that for each energy level we get the complete Maxwell-Boltzmann function, and we have represented it in the third column of Figure \ref{graphs}. Let us repeat that these are the exact values of the Maxwell-Boltzmann function, to which therefore (\ref{maxwellboltzmanndistribution}) is a continuous approximation. We have also attached a graph of this Maxwell-Boltzmann function for the situation of four words distributed over seven energy levels with the total energy equal to seven, it is the red graph shown in Figure \ref{Bose-Boltzmann-FermiSevenFourEnergyLevelsWordsfigure}. Before we further analyze the details of this exact version of the Maxwell-Boltzmann function, let us explain how we obtain the corresponding exact version of the Bose-Einstein function and the corresponding exact version of the Fermi-Dirac function.
\begin{table}[h!]
\centering
\caption{The calculated data for the Bose-Einstein, Maxwell-Boltzmann and Fermi-Dirac functions belonging to the example of `four words distributed over seven energy levels with total energy equal to seven', whose raw data and previous calculated data can be found in Table \ref{fourwordssevenenergylevels}.The graphs of these data can be seen in Figure \ref{Bose-Boltzmann-FermiSevenFourEnergyLevelsWordsfigure}, Figure \ref{Bose-Boltzmann-SevenFourEnergyLevelsWordsLogfigure} and Figure \ref{Bose-Boltzmann-FermiEnergiesSevenFourEnergyLevelsWordsfigure}.}
\label{graphs}
\begin{tabular}{p{2.4cm}p{0.9cm}p{0.9cm}p{0.9cm}p{0.9cm}p{0.9cm}p{0.9cm}p{0.9cm}p{0.9cm}p{0.9cm}p{0.9cm}p{0.9cm}}                         
\hline 
\hline                             
\multicolumn{12}{c}{Graphs} \\
\hline
\hline
Graphs			&	B-E	&	M-B	&	F-D  	&	Log  	&	 		&			&			&	  		&Energies &	 &	\\
\hline
0	& 1.18 		& 1.2  		& 1.25  		& 	&  		&  		& 			&  		& 0 		& 0		& 0 \\
1					& 1 		& 0.93  & 0.88 & 0		& 0 		& -0.03	& 	& 1 		& 1		& 0.93 	& 0.82 \\
2					& 0.72 	& 0.7   	& 0.63 & 0.30 	& -0.14& -0.15 	& 	& 2 		& 1.45 	& 1.02 	& 0.64 \\
3					& 0.45 	& 0.5  	& 0.63 & 0.48 	& -0.34& -0.30 	&	& 3 		& 1.36 	& 0.68 	& 0.43\\
4					& 0.27 	& 0.33  & 0.25 & 0.60 	& -0.56& -0.48 	&	& 4		& 1.09 	& 0.36 	& 0.09 \\
5					& 0.18	& 0.2  	& 0.25 & 0.70 	& -0.74& -0.70 	& 	& 5 		& 0.90 	& 0.18 	& 0.05 \\
6					& 0.09 	& 0.1	& 0.13 & 0.78 	& -1.04& -1  	& 	& 6  		& 0.55 	& 0.05 	& 0.006 \\
7					& 0.09	& 0.03  & 0    & 0.85 	& -1.04& -1.48 & 			& 7  		& 0.64 	& 0.02 	&  \\
\hline                  
\end{tabular}
\end{table}

For that, we go back to Table \ref{fourwordssevenenergylevels}, the third part of this table, under the heading Bose-Einstein. What we called the Maxwell-Boltzmann weights in the second part of Table \ref{fourwordssevenenergylevels}, represented on the row `M-B weights' of the second part of Table \ref{fourwordssevenenergylevels}, now become the Bose-Einstein weights, represented on the row `B-E weights' of the third part of Table \ref{fourwordssevenenergylevels}. These are all equal to $1$ divided by the number of macro states, i.e. $1/11$. We thus encounter here the transformation of the Maxwell-Boltzmann weights which are naturally given by the presence of in principle distinguishable micro states, to Bose-Einstein weights which are all taken equal per macro state. It is this transformation that Einstein found highly inconvenient and which he wished to attribute to the presence of a still entirely mysterious force that causes photons to clump together in the same states. We have explained above how in human cognition and human language the force of `meaning' can play this role, and in analyzing our example further we will show this more concretely. The remainder of the calculations proceed totally identically to how we explained it in detail for the Maxwell-Boltzmann case, only that this time we substitute the Bose-Einstein weights for the earlier Maxwell-Boltzmann weights. We thus obtain the values for the exact Bose-Einstein function, also in an identical way as we did for the values of the exact Maxwell-Boltzmann function, and we find them in Table \ref{graphs}, in the second column and its representation in a graph in Figure \ref{Bose-Boltzmann-FermiSevenFourEnergyLevelsWordsfigure}, namely the blue graph. 
At this point it becomes interesting to compare these graphs, the red and blue graphs of Figure \ref{Bose-Boltzmann-FermiSevenFourEnergyLevelsWordsfigure} with the graphs of the Bose-Einstein and Maxwell-Boltzmann functions of the Winnie the Pooh story, i.e. the red and green graphs of Figure \ref{piglethaffalunmpgraphfigure}. To allow for a more accurate comparison, we also constructed the graphs of the log/log functions, whose values can be found in Table \ref{graphs}, the fifth, sixth and seventh columns, which we can then compare with the log/log functions of the Bose-Einstein functions and the Maxwell-Boltzmann functions of the Winnie the Pooh story, more specifically depicted in the red and green graphs of Figure \ref{piglethaffalunmploggraphfigure}. 
Although our example of four words distributed over seven energy levels so that the total energy is equal to seven is an extremely small example, we already see similarities with the graphs of the continuous functions. The Bose-Einstein graph extends higher than the Maxwell-Boltzmann graph in $0$ and is less curved in the log/log version, as is also the case for the Bose-Einstein and the Maxwell-Boltzmann of the Winnie the Pooh story. The higher value at $0$ is associated with the tendency for words to accumulate at small energies when there is not much energy available, ultimately leading to the formation of a Bose-Einstein condensate with all words in the energy level $0$. Before showing how the identification of a `meaning dynamics' gives us a finer understanding of the differences between Bose-Einstein and Maxwell-Boltzmann, we wish to elaborate on the Fermi-Dirac version of this situation. 
\begin{table}
\centering
\caption{The `meaning dynamics' belonging to the example of `four sentences distributed over seven energy levels with total energy equal to seven'. Both for the B-E actions and for the F-D actions a number greater than 1 means that an attraction takes place, and a number less than 1 means that a repulsion takes place.}
\label{meaningdynamics}
\begin{tabular}{p{2.4cm}p{0.9cm}p{0.9cm}p{0.9cm}p{0.9cm}p{0.9cm}p{0.9cm}p{0.9cm}p{0.9cm}p{0.9cm}p{0.9cm}p{0.9cm}}
\\\hline
\multicolumn{12}{c}{Meaning Dynamics} \\
\hline
					&  rep 	&  att	& att		&  att	& att		& rep 	& rep 	& att		& att		& att		& rep \\
F-D action		& 0 		& 1.25 	& 1.25 	& 1.25 	& 1.25 	& 0.625&0		& 1.25 	& 1.25	& 1.25	&0\\
F-D numbers& 0		& 15   	& 15    	& 15 	& 15		& 15 	& 0   	& 15 	& 15 	& 15		& 0\\
					&  att 	&  rep	& rep	& rep	& rep	& rep 	& att		& rep	& rep	& rep	& rep \\
B-E action		& 2.73	& 0.91  & 0.91 	& 0.91 	& 0.91 	& 0.45 	& 2.73 	& 0.91  & 0.91 	& 0.91  & 2.73\\
B-E numbers& 10.9 	& 10.9  & 10.9  & 10.9 & 10.9 	& 10.9 	& 10.9  & 10.9 	& 10.9 	& 10.9 	& 10.9 \\
M-B numbers& 4 		& 12   	& 12    	& 12   	& 12 	& 24 	& 4 		& 12    	& 12   	& 12 	&4 \\
\hline
\end{tabular}
\end{table}

We can introduce the Fermi-Dirac case by again assigning different weights to the macro states, using our knowledge of the Pauli exclusion principle. In modern quantum mechanics, both Bose-Einstein and Fermi-Dirac are  mainly mathematically derived, by applying an exchange symmetry to the Hilbert space, but we are not following this path for the time being. Rather, we will simply apply the Pauli exclusion principle, and assume that words cannot be in the same state when these words are the elements of a memory. In a memory there should indeed be very careful avoidance of memory elements to remain distinguishable. Computer memories, which we all use daily, do not allow, for example, the same name to be given to two different files. On the other hand, we will also assume that there are internal parameters in that memory that classify different external equal states as different within an internal space. 
In analogy to the internal parameter space possessed by a quantum particle through its spin, spin up and spin down, we will also assume for our example of four words distributed over seven energy levels that there are two places available per energy level and per macro state in terms of internal space.
Hence, only those macro states are allowed that contain no more than two words at a given energy level, one of those words is then classified in memory in one of these places, and the other word in the other places. Being excluded as a consequence of the Pauli exclusion principle translates into possessing Fermi-Dirac weight $0$, and thus that applies concretely to macro states $1$, $7$ and $11$, all other macro states are admitted to the memory realm of Fermi-Dirac. We can see this Fermi-Dirac situation worked out in the third part of Table \ref{fourwordssevenenergylevels}, since there are eight macro states left in the Fermi-Dirac memory realm, each of them is given a weight of $1/8=0.125$, and in the second row of the third part of Table \ref{fourwordssevenenergylevels} under the heading `Fermi-Dirac' these `Fermi-Dirac weights' can be found. To then calculate the Fermi-Dirac function we proceed identically as we did for the Maxwell-Boltzmann function, except that we use the Fermi-Dirac weights in place of the Maxwell-Boltzmann weights. The values of the exact Fermi-Dirac function that we thus obtain can be found in the fourth column of Table \ref{graphs}, and the graph of this function can be found in Figure \ref{Bose-Boltzmann-FermiSevenFourEnergyLevelsWordsfigure}, it is the green graph. Let us consider Figure \ref{Bose-Boltzmann-FermiSevenFourEnergyLevelsWordsfigure} for a moment. As we already noted, the red graph represents the exact Maxwell-Boltzmann function of our example, the blue graph the exact Bose-Einstein function, and the green graph the exact Fermi-Dirac function. If one considers the three graphs, the blue one, the Bose-Einstein, and the green one, the Fermi-Dirac, seem to move around the red graph, the Maxwell-Boltzmann, in an opposite way. This is no coincidence and is connected to the `meaning dynamics' that we now wish to discuss and analyze.

To study the `meaning dynamics', we compiled Table \ref{meaningdynamics}. We put ourselves in the typical picture of statistical thermodynamics, as we already did when we discussed the micro states and the quantities of these micro states per macro state. Each micro state has an equal chance of realizing itself and then the Maxwell-Boltzmann numbers represent the weights by which the macro states will realize themselves. We call these numbers the M-B numbers, calculated them already, and they can be found again in the eighth row of Table \ref{meaningdynamics}. The sum of all these numbers is equal to $120$, and we already mentioned that this is the total number of micro states for the situation considered by us of four words distributed over seven energy levels with total energy equal to seven. We now imagine that the Bose-Einstein statistics emerge as a consequence of the presence of a `meaning force', the mysterious force that Einstein hypothesized to be present and to cause photons to clump together in the same states. Bose-Einstein statistics gives equal weight to each macro state, which means that the total number of micro states are distributed proportionally among the macro states under a Bose-Einstein regime. That is how we determine the Bose-Einstein numbers, found in the seventh row of Table \ref{meaningdynamics}. If for a macro state its B-E number is greater than its M-B number, then this means that an `attractive meaning dynamics' takes place whose magnitude we can calculate by dividing both numbers by each other, and we call this division the B-E action. If the B-E action is greater than $1$ there is attraction, and if the B-E action is less than $1$ there is repulsion. In the sixth row of Table \ref{meaningdynamics}, the B-E actions of the different macro states can be read, there are three attractions, the first, sixth and eleventh macro state, and the remaining eight macro states are repulsions. By looking more closely at the attractions and repulsions, we see that indeed it is a clumping together toward more equal states. If we do a similar calculation for Fermi-Dirac, we see that the attractions are generally those macro states that are repulsions for Bose-Einstein. But this is not always the case, for example macro state number $6$ is a repulsion for both Bose-Einstein and Fermi-Dirac. This means that the `meaning dynamics' of Bose-Einstein is the underlying dominant dynamics, and Fermi-Dirac is also determined by that underlying dynamics. Only for Fermi-Dirac there is the additional Pauli exclusion principle causing that reversal for some macro states.

Before explaining further how we consider the Fermi-Dirac situation, we wish to formulate a brief remark in between. Thinking about the presence of a `meaning dynamics', this perhaps explains why quantum structures have emerged in a fruitful way in a research area such as `information retrieval', where one is very focused on `meaning in language', or what is called `semantics'. Let us now reason further about the Fermi-Dirac case, or why introducing internal variables for the memory realm in which words can nest, and we are thinking very specifically of human memory this time, makes sense. 
Before we get to this point, however, let us introduce the basic elements of how we look at words, concepts, sentences, and human language in general, which originated in the period when one of us was studying concepts, with a focus on the well-known `guppy effect in concept theory', a study that then contributed to the emergence of quantum cognition \citep{gaboraaerts2002,aertsgabora2005a,aertsgabora2005b}. We consider a word or a concept as an entity that can be in different states and thus introduce our interpretation of what in concept research is called `prototype theory', originally introduced by Eleanor Rosch \citep{rosch1978,rosch1983}. We explicitly introduce the notion of `state of a concept', where this is done implicitly in the traditional version of prototype theory by means of the properties. The concept {\it Cat} is a state of the concept {\it Animal}, so an `exemplar' is a `state' of the concept of which it is an exemplar. However, that is not the only way in which states of concepts form. Likewise, when the word `animal' is in a text, the presence of the text surrounding the word `animal' changes the state of the corresponding concept {\it Animal}. Exemplars of a concept are states of that concept, but also `contexts that surround a concept in a text' determine states of that concept. Not only words but also symbols can indicate concepts, which is for example often the case in mathematics, hence the notion of `concept' is broader than the notion of `word'.
Note in that sense that in \citet{aertsbeltran2020} we introduced the new notion of `cogniton', to denote a `quantum of human language'. Often presented as rival theory, in addition to Rosch's protoype theory there also exists what is called the `exemplar theory' in concept theory \citep{nosofsky1988,nosofsky1989}. Let us outline how we want to further elaborate the Fermi-Dirac part for human cognition, and why thereby choosing the macro states for the Fermi-Dirac case, as we did in the example of the four words with seven energy levels, makes sense as a simple example. The basic idea is that the exemplars stored in human memory, and lending a helping hand in categorizing according to exemplar theory, give way to the internal parameters for the states in which the same concepts can nevertheless be present in human memory in different states, thus opening the way for the presence of a Fermi-Dirac statistics for human memory. More concretely, the concept {\it Cat}, in its bosonic nature is one and unique, and may well, since bosons allow it, be in this one and the same state in multiplicity, for example {\it Five Cats}. 
\begin{figure}[h!]
\begin{center}
\includegraphics[width=10cm]{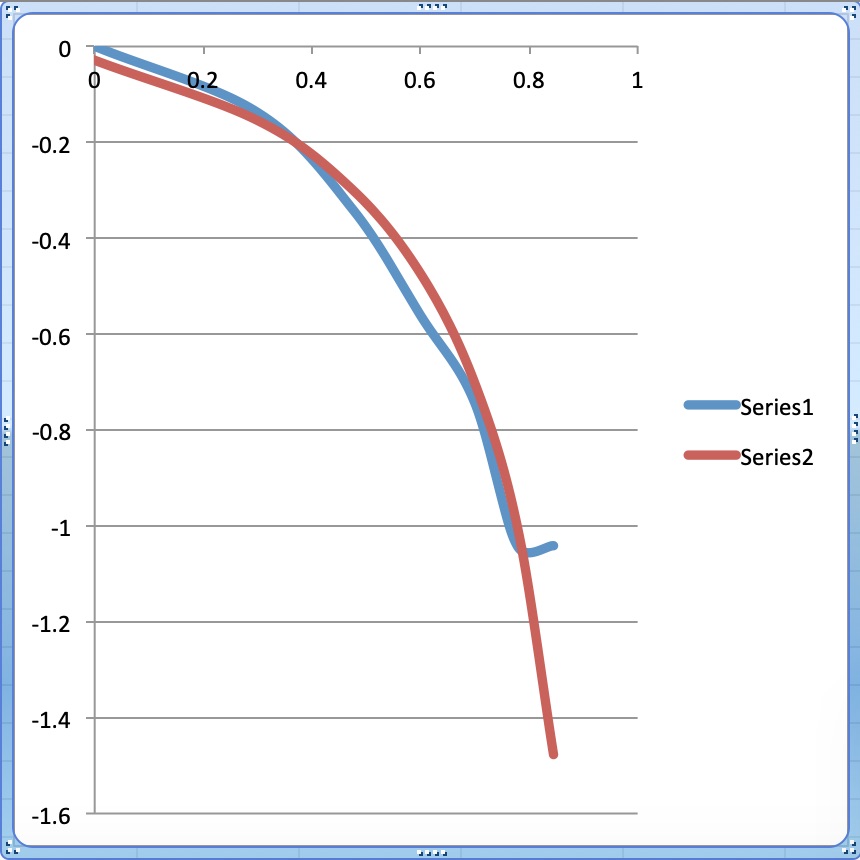}
\end{center}
\caption{The blue and red curves are the log/log graphs of the Bose-Einstein and Maxwell-Boltzmann functions of the situation of four words distributed over seven energy levels with total energy equal to seven.}
\label{Bose-Boltzmann-SevenFourEnergyLevelsWordsLogfigure}
\end{figure}
That same concept {\it Cat}, in the realm of human memory, can be located in different states, in, as it were, an internal parameter space, and these are the concrete exemplars of cats that a particular human being has stored in his or her memory from his or her personal experience.
Let us additionally repeat the reasoning from a previous article of which one of us was a co-author, because from a slightly different angle it also illustrates the difference between bosonic and fermionic as far as human cognition is concerned \citep{aertssozzo2015}. We consider the situation of a group of our ancestors collaborating during the activity of `hunting', and assume that there are several of them who always, or at least very regularly, go hunting together. Each of them will carry a series of conceptual representations of what hunting is, in his or her memory, with specific exemplar situations that are important from a personal perspective. 
This presupposes a multitude of memory states, all important to what hunting is, and this set of states is stored in what we called an internal parameter space. The structure of this state space is fermionic, only one exemplar element is admitted to a state, otherwise the state space could not function like a memory. But in communicating with the others with whom regular hunting takes place, it is necessary to make the notion {\it Hunting} independent of this personal internal memory structure. And that is how the bosonic concept {\it Hunting} makes its appearance within the communication between the hunters and finds a place in their language. Here, within language, it is a bosonic concept, where a multiplicity can clump together. It makes sense to speak of {\it Several Huntings} and within this interpersonal language they are all `identical' and `indistinguishable'. 
We already mentioned above the spin of a quantum entity as an example of how quantum entities carry such an internal parameter space. But also the different orbits within the internal space of an atom on which electrons can find themselves can be counted among such internal parameter spaces, for these orbits also allow electrons to be in multiples, albeit different states, within an atom, but only two electron can be in each orbital place, one with spin up and another one with spin down. 
\begin{figure}[h!]
\begin{center}
\includegraphics[width=10cm]{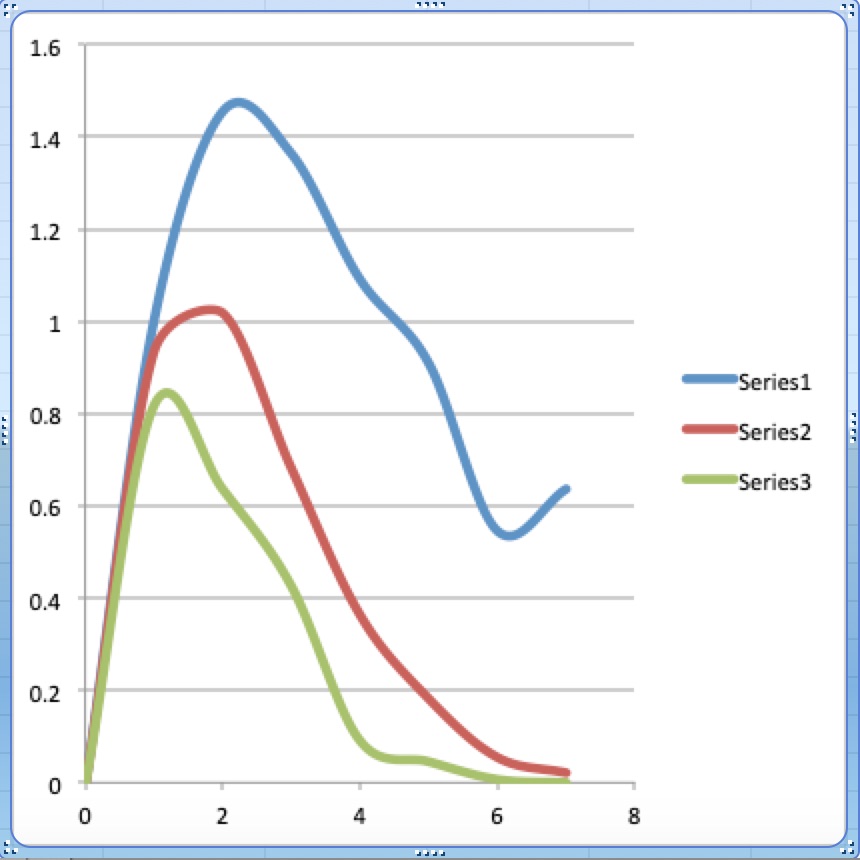}
\end{center}
\caption{
The blue, red and green curves are the Bose-Einstein energy radiation function the Maxwell-Boltzmann energy radiation function and the Fermi-Dirac energy radiation function of the situation of four words distributed over seven energy levels with total energy equal to seven.}
\label{Bose-Boltzmann-FermiEnergiesSevenFourEnergyLevelsWordsfigure}
\end{figure}
A very simple memory structure in terms of human language is a piece of paper on which a story is written. The words on the piece of paper are clearly embedded in a fermionic structure, which is so simple that there can only be one word in one place. Thus, there are no internal parameter spaces in this simple memory structure to allow for more states within the fermionic structure. Unless one allows punctuation or other annotations to the text to play this role. Highlighting a piece of text, for example, puts this piece in a different state, than before the piece was highlighted. Accents can also change the state of a word within the very limited memory space of a piece of paper and a written narrative. 

We want to end this Section \ref{meaningdynamics} by using a result that we elaborated in a recent publication related to the presence of entanglement in the combination of words in a text because it also sheds light on our analysis of `meaning dynamics' in this Section \ref{meaningdynamics} and the difference between bosons and fermions and language and memory \citep{aertsbeltran2022}. We have identified entanglement in human cognition and language and studied various aspects of it in our Brussels research group over the years \citep{aertssozzo2011,aertssozzo2014,aertsetal2018b,aertsetal2018c,aertsarguelles2018,aertsetal2019a,aertsetal2021b}. In \citet{aertsbeltran2022} we reflected on all these examples of entanglement together, and showed how in a text of a story entanglement is essentially important in order to make the formulation of the story as precise as possible in terms of content and meaning. We demonstrated this by introducing the von Neumann entropy into human cognition and language, and examined how it is precisely the presence of entanglement that makes the von Neumann entropy of the entire text of the story smaller than the von Neumann entropy of separate words in the text of the story. In this sense, then, we can speak of a `decrease in entropy' and hence an `increase in order' for the entire text, as a consequence of entanglement, relative to the order grasped by individual words. These findings made us think further, and we now believe that it is possible to understand and explain the process we identified in \citet{aertsbeltran2022}, from a fundamental analysis about what concepts represent in relation to human experience. Since this new insight about entanglement can also contribute, as we will explain, to a better understanding of what the `meaning dynamics' represents that we introduced and studied in this Section \ref{meaningdynamics}, we bring here, as the concluding part of this Section \ref{meaningdynamics}, a description of this insight about entanglement. 
We begin our explanation with a thought experiment. Suppose we compose a text in the following way, we take a dictionary, and install a system to choose words at random from the dictionary. We can use dice to first choose a page of the dictionary and then a random word on that page, or any other more sophisticated way, it is not important for our thought experiment how exactly this random choice takes place, as long as it is random. In this way, we choose a first word, then a second, then a third, and so on, until we have a text of a certain size similar to the texts of stories we analyzed in \citet{aertsbeltran2020}. Everyone understands that even for a much smaller text of a few paragraphs, a text composed in such a way does not contain content and meaning. But is that really the case? We need to express more precisely what is going on with such a composite text. The writer, or composer, of the text cannot put meaning or content into such a composite text. A reader of this text can, and often will, do so with success, at least if the text is not too long. In a precisely formulated text, however, there is content and meaning present along both sides, introduced by the writer, intentionally or less intentionally, and introduced by the reader when reading, intentionally or less intentionally. What we are expressing here is that human minds are entities that wish to experience `meaning', even in experiences that are constructed in a way that, in themselves at least, they carry no meaning. That human minds are able to turn a text constructed in the random manner explained in our thought experiment into content and meaning `after all', and if the text is not too long this can even be done without too much effort, is at least partly because `words' already carry meaning, by definition of what they are and how they came about. So the random process of our thought experiment is only a partial destruction of meaning, because we continue to build the text with building blocks that individually still carry meaning. These building blocks, that is, the words, came into existence by giving the most frequent and important experiences `a name'. But, calling them `building blocks of a story' probably already contributes to later misunderstandings, because it uses an analogy, like `bricks are building blocks of a house' for example. Originally, probably, words were combined to give a name to experiences that were not frequent and/or important enough to be `assigned one word'. In part this is also at least due to the fact that there were many more experiences to name than it was possible to form words, consisting of sounds, and later of letters. If it is true, that the above describes, in an admittedly simplified way, how language arose, then entanglement must be used in combining words into an expression that stands as a substitute for 'one word'. Many other creative uses have arisen in parallel with this basic use, but that does not prevent the presence of entanglement from remaining an essential part of combining words. Let us also note, and it would lead us too far to elaborate on that here, but many readers certainly know enough the mathematics of quantum mechanics to understand this remark, that in a very fundamental way the mathematical formalism of quantum mechanics in complex Hilbert spaces incorporates this composition structure. Two quantum entities are described in the tensor product of their respective Hilbert spaces, and that tensor product is again a Hilbert space with as dimension the product of the dimensions of the Hilbert spaces of the sub entities. Mathematically, this means that two quantum entities when they combine become `one quantum entity', and indeed it is the non-product states of that combined unified quantum entity that bring about the presence of entanglement. On the other hand, it is also a confirmation of the research that one of us did, many years ago, when the tests of the Bell inequalities were still in progress, and where in this research it was shown that the mathematical formalism of quantum mechanics is purely structurally incapable of describing separated quantum entities. Indeed, there are two quantum axioms of the axiomatic mathematical structure of this quantum formalism that stand in the way of this \citep{aerts1982}. It is not our intention to elaborate on this now, but in future work we certainly want to explore this situation in greater depth again, especially now that many other aspects of this situation have become clear in the meantime. Indeed, from the analysis of Bose-Einstein and Maxwell-Boltzmann that we made here, it appears that separated or not separated entities will also determine the choice between these two, Maxwell-Boltzmann or Bose-Einstein (or Fermi-Dirac). However, consider again our example of the two children who choose a cat and a dog, but who, after they could not come to a choice together, are asked to choose separately. It remains the case that at that point Maxwell-Boltzmann is valid, and we think that even for quantum entities the context may well be such that `separated' and `Maxwell-Boltzmann' are the ones that are valid. Still it has not become clear how exactly this kind of context looks like in all details in physical reality. If we consider the examples of the combinations of concepts with which we violated the Bell inequalities, so that the presence of entanglement was demonstrated, it was always a `meaning connection' that was at the root of this violation of the Bell inequalities, and that gave rise to the correlations responsible for it \citep{aertssozzo2011,aertssozzo2014,aertsetal2018b,aertsetal2018c,aertsarguelles2018,aertsetal2019a,aertsetal2021b}. We think it is worthwhile to investigate even more thoroughly how the `substance of meaning' works with respect to whether or not different kinds of contexts are allowed to realize themselves and thus try to understand even better the manifestation or not of `separated' and `Maxwell-Boltzmann' or the forcing or not of `entanglement' and `Bose-Einstein'.   

\section{Conclusions} 
We have continued our search in this article for an explanation of our experimental findings of the presence of the Bose-Einstein statistics in human cognition and language. We have outlined the genesis of these findings and how they have recently culminated in a very convincing practical watertight proof of the Bose-Einstein statistical structure of texts of stories. With it, we want to show that quantum cognition can broaden in the application of quantum structures in human cognition, by giving, in addition to the vector spaces and probability models of quantum mechanics, those aspects of quantum mechanics associated with identity, indistinguishability, and quantization their place in human cognition and language. More specifically, we argue that words can be considered as quanta of human cognition, and that the notions of `energy' and `energy level', and even more, a radiation law scheme, can be fruitfully introduced into human cognition, in order to analyze from there, based on such a scheme as ground, human thought, human communication, human memory, and language. The occurrence of the systematics of Zipf's and Pareto's laws in the many areas of human society, language, culture, and economics, and the demonstration of our proposed `radiation scheme' as a theoretical underpinning of Zipf's and Pareto's laws, we cite as arguments in support of our proposal. That our proposed radiation and quantization scheme provides a theoretical foundation for Zipf's and Pareto's laws promises that this scheme has broad applicability in many areas of science, more specifically and among others, in psychology and economics. The identification of a `meaning dynamics' underlying the presence of the Bose-Einstein statistics suggests that this broad applicability is likely linked to human meaning-making as a contributing cause of many of these situations, an intriguing possibility that we intend to explore further in depth. We elaborated on the historical events in the period of the Old Quantum Theory, and the discussions between Planck, Einstein and Ehrenfest, with Bose briefly appearing on the scene, to better understand the genesis of the Bose-Einstein statistics in quantum mechanics. But there is another reason why we enthusiastically mentioned these historical elements with due detail. Indeed, it turned out that the insistence of the protagonists, and especially of Einstein, on structural similarities between the radiation of light, on the one hand, and the behavior of a gas of bosons, on the other, against superficial obstacles, brought out the depths of quantum mechanics. The torch of this kind of persistence of a theoretical and even aesthetic nature was later picked up by Schr\"odinger, Heisenberg, Dirac and others. By citing these historical elements, we wanted to highlight an analogy with quantum cognition, and thus give heart to researchers in quantum cognition. As we already noted, for us the research in quantum cognition is twofold. On the one hand, it aims to better understand and describe human cognition with the help of models originating in quantum mechanics. On the other hand, and certainly in our view no less important, it is also the intention to better understand quantum mechanics, and thus the physical world, as a consequence of a pertinent application of its structures in human cognition. In that sense, the following might also have been a title for this article, `A human cognition and language model for the radiation and quantization of light'.
One of the questions that, thinking also about this second component of our research in quantum cognition, indeed now concerns us is the following. `What is the equivalent in the physical world of the meaning dynamics that we identified as underlying to the presence of the Bose-Einstein statistics in human cognition?'; or in the words of Albert Einstein, `what is that mysterious force that causes bosons to clump together into the same states?' Although this was not the focus of our discourse, the insights we have set forth in the foregoing can likewise serve as evidence for the conceptuality interpretation of quantum mechanics we are working on \citep{aerts2009b,aerts2010,aertsetal2018d,aertsetal2019b}, which indeed starts from the basic idea that quantum particles are not objects but concepts. 

\bigskip
\noindent
{\bf Acknowledgments}

\noindent
This work was supported by QUARTZ (Quantum Information Access and Retrieval Theory), the Marie Sklodowska-Curie Innovative Training Network 721321 of the European Unions Horizon 2020 research and innovation program.
 
\bigskip
\noindent
{\bf Data Availability Statement}

\noindent
The datasets for Table \ref{piglethaffalunmp} can be found at the link \url{https://www.dropbox.com/s/xshgz2xmaqhtz4e/WinniethePoohData.xlsx?dl=0}, and the data set for Figure \ref{BlackBodychartfigure} can be found at the link \url{https://www.dropbox.com/s/70rxque66fss2ze/BlackBodyGraphSOURCE.xlsx?dl=0}

\end{document}